# A validated multi-agent simulation test bed to evaluate congestion pricing policies on population segments by time-of-day in New York City

**Brian Yueshuai He, Jinkai Zhou, Ziyi Ma, Ding Wang, Di Sha, Mina Lee, Joseph Y. J. Chow [*], Kaan Ozbay**

C2SMART University Transportation Center, New York University Tandon School of Engineering, Brooklyn, NY, USA

[*]Corresponding author email: joseph.chow@nyu.edu

## Abstract

Evaluation of the demand for emerging transportation technologies and policies can vary by time of day due to spillbacks on roadways, rescheduling of travelers' activity patterns, and shifting to other modes that affect the level of congestion. These effects are not well-captured with static travel demand models. We calibrate and validate the first open-source multi-agent simulation model for New York City, called MATSim-NYC, to support agencies in evaluating policies such as congestion pricing. The simulation-based virtual test bed is loaded with an 8M+ synthetic 2016 population calibrated in a prior study. The road network is calibrated to INRIX speed data and average annual daily traffic for a screenline along the East River crossings, resulting in average speed differences of 7.2% on freeways and 17.1% on arterials, leading to average difference of +1.8% from the East River screenline. Validation against transit stations shows an 8% difference from observed counts and median difference of 29% for select road link counts. The model is used to evaluate a congestion pricing plan proposed by the Regional Plan Association and suggests a much higher (127K) car trip reduction compared to their report (59K). The pricing policy would impact the population segment making trips within Manhattan differently from the population segment of trips outside Manhattan: benefits from congestion reduction benefit the former by about 110%+ more than the latter. The multiagent simulation can show that 37.3% of the Manhattan segment would be negatively impacted by the pricing compared to 39.9% of the non-Manhattan segment, which has implications for redistribution of congestion pricing revenues. The citywide travel consumer surplus decreases when the congestion pricing goes up from $9.18 to $14 both ways even as it increases for the Charging-related population segment. This implies that increasing pricing from $9.18 to $14 benefits Manhattanites at the expense of the rest of the city.

# 1. Introduction

In the era of the Internet of Things (IoT), cities are facing a growth in new technologies and operational models in the context of "smart cities." A fine example of the impact this paradigm shift has on mobility options is shown in Figure 1 under a Mobility-as-a-Service (MaaS) paradigm. Whereas traditional transportation planning tools focus on evaluation of roadway infrastructure and public transit alternatives, emerging mobility services and technologies play a much bigger role today (Chow, 2018).

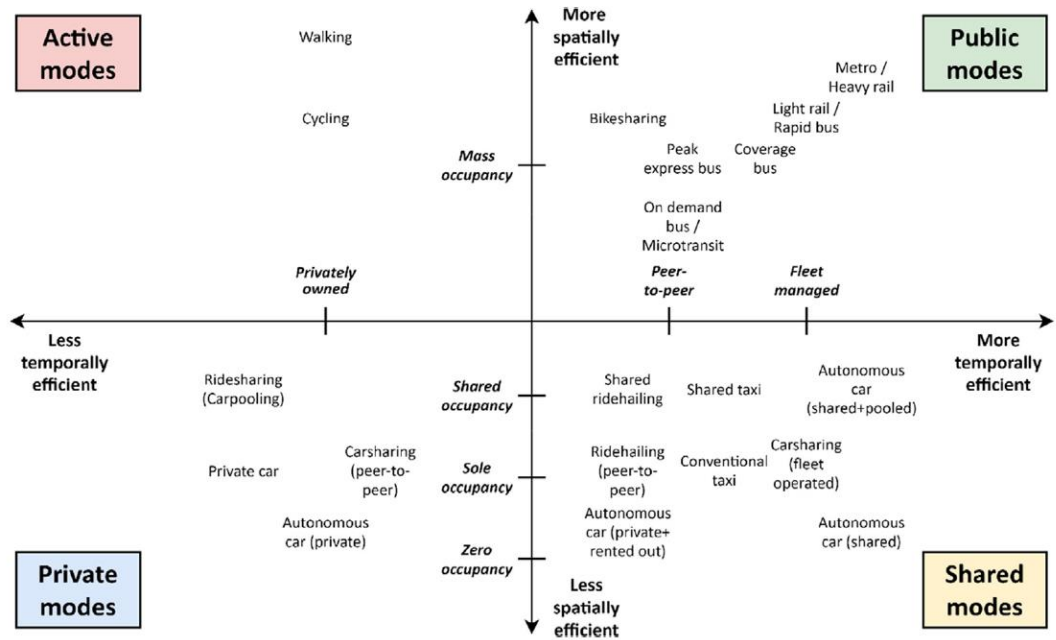

Figure 1. Spectrum of modes available in a MaaS paradigm (source: Wong et al., 2020).

To grapple with these emerging technologies, city agencies need to evaluate operational scenarios imposed by the private sector (e.g., what is the impact of e-hail ride-sourcing on traffic congestion?) or when considering what-if scenarios related to new operating policies. This is especially important because technologies companies developing public products need to gain approval from public agencies before they can deploy in that region. As such, public agencies need to evaluate the product's impact on the community.

How should government agencies test such products? For most engineered products, the evaluation and testing phases of a product out of research and development is either prototyping or deployment testing in the field. However, transportation products like policies and operating technologies face different challenges than conventional technologies because of their public nature. Transportation technologies deployed to the field can be both financially and socially costly, as unproven technologies may end up costing lives if something goes wrong. Furthermore, even a successful deployment in one city may not be indicative that the same technology can work well in another city because each city is a different market.

Prototyping serves to verify that a technology works, but it does not consider how the technology may impact a community considering the behavior of its population. This gap in the innovation process for transportation technologies suggests a need for a deployment testing framework that falls between prototyping and field piloting. The lack of a consistent deployment modeling and testing phase between prototyping and deployment pilot can lead to higher failure rates in emerging technologies. Companies like Chariot, Bridj, ReachNow, Bird, among others (see Chow, 2018, and Chow et al., 2020a, for other examples), have all failed to operate sustainably. An *ex post* analysis of Kutsuplus microtransit suggests the operating conditions might not have been adequate to maintain such services (Haglund et al., 2019).

A new mobility provider needs approval from city officials if they plan to enter the market. However, the officials in New York City (NYC) have limited policy tools or models to evaluate emerging technologies. One is the New York Best Practice Model (NYBPM), developed by the New York Metropolitan Transport Council (NYMTC). The NYBPM is a regional travel demand model to forecast travel patterns in the NYMTC region. It covers 28 counties of New York, New Jersey and Connecticut and both the road and public transit networks are incorporated. The latest released version does not capture the demand for for-hire vehicles (FHVs) like Uber, Lyft, Via, and other transportation network companies (TNCs) since 2015. The NYBPM is designed for long term capital planning rather than for quick response evaluation of operating policies introduced by emerging transportation technologies or circumstances which are often dynamic and impact travelers' preferences throughout the day.

Another tool is the Balanced Transportation Analyzer (BTA) developed by the Nurture Nature Foundation (NNF). The BTA is an intricate spreadsheet model to help analyze the impacts of transportation fares and other variables. The Regional Planning Association (RPA) published a report about the congestion pricing analysis in Manhattan using this model (RPA, 2019). However, this model does not capture spatiotemporal dependencies within the city.

Congestion pricing, algorithms for micro-transit or bikeshare rebalancing, electric vehicle fleets with dynamic fast charging activities, and pandemics like COVID-19 all have one thing in common: they impact travelers' choices throughout the day, which in turn impact the dynamics of traffic congestion throughout the day. Traffic dynamics are currently not accounted for in any existing policy tools in NYC, nor in most cities around the world. For example, the BPM has a complex microsimulation of population travel choices, but the network assignment is determined using a static traffic assignment model via TransCAD. This means, for example, it would not capture traffic spillbacks during a peak hour and the effect they have on making travelers depart earlier. As such, congestion pricing models based on TransCAD (e.g. Baghestani et al., 2020) are not sensitive to time of day effects of congestion pricing on travelers.

The urgency for NYC is paramount: on top of the congestion pricing policy to be implemented and new emerging modes like micromobility and large-scale public transit projects (e.g. Brooklyn-Queens Connector (BQX) and bus network redesigns), there is now the COVID-19 pandemic and new policies regulating micromobility and urban cargo bike deliveries. Yet, there is no citywide policy tool available to agencies to evaluate these scenarios. There is a need to develop such a policy tool for NYC that is: 1) sensitive to traffic dynamics (at least at the mesoscopic level) as they relate to travelers' activity-based choices throughout the day, and 2) capable of incorporating multiple travel modes that include modes like micromobility and FHVs.

The contribution of this study is the first validated agent-based simulation-based virtual test bed for NYC to fulfill this need. The test bed is built using an open-source platform called Multi-Agent Transport Simulation (MATSim) and makes use of a synthetic population developed by He et al. (2020). The validated test bed is applied to a case study to demonstrate how such a tool can offer insights that existing policy tools cannot: a cordon-based congestion pricing plan as proposed by the RPA. The scenario is the first in the literature to use multi-agent simulation to provide new insights to NYC policymakers. Other applications can also be found in related studies for COVID-19 (Wang et al., 2020), the BQX (Chow et al., 2020b) and Brooklyn bus network redesign (Chow et al., 2020c).

The rest of this study is organized as follows. Section 2 reviews the literature of agent-based simulation and gives an overview of the MATSim platform. Studies conducted to evaluate scenarios in NYC are also reviewed. The data used for the policy tool and case studies are introduced in Section 3. Section 4 covers the calibration methodology and the validation results. Section 5 presents the analysis and discussion of the congestion pricing and BQX scenarios. Section 6 concludes the study.

## 2. Literature review

First, we review the research of the ABMS in transportation area and justify the model that we end up with. Then, an overview of the simulation platform MATSim is provided. Studies focusing on analyzing the impacts of city-scale policies are reviewed, especially in the context of NYC.

**2.1 Agent-based modeling and simulation**

Agent-Based Modeling and Simulation (ABMS) (von Neumann, 1966; Bonabeau, 2002) can model complex heterogeneous agents with interaction rules and agent learning. ABMS has been applied to many problems in the transportation area (see Dia, 2002; Hidas, 2002; Zhang, 2006; Rieser et al., 2016). Macal and North (2006) classified the applications of ABMS into two categories: "Small, elegant, minimalist models" and "Large-scale decision-support systems". The latter one is more suitable to facilitate the emerging needs of policymakers. Djavadian and Chow (2017a, b) demonstrate how agent-based simulation can capture market equilibration for dynamic transportation systems. The framework is shown to reach a stochastic user equilibrium when populations are sampled sufficiently (Djavadian and Chow, 2017a), which provides a basis for agent-based simulations of such transportation systems.

There are several well-known ABMS platforms designed to support decision-making, including but not limited to Transportation Analysis and Simulation System (TRANSIMS) (Nagel et al., 1999), MATSim (Balmer et al., 2009), Sacramento Activity-Based Travel Demand Simulation Model (SACSIM) (Bradley et al., 2012) Simulator of Activities, Greenhouse Emissions, Networks, and Travel (SimAGENT) (Goulias et al., 2011), Polaris (Auld et al., 2016), SimMobility (e.g., Nahmias-Biran et al., 2019), etc. TRANSIMS was a first-generation tool developed by the Federal Highway Administration (FHWA), after which the creators used to produce the next generation tool MATSim.

**2.2 Overview of MATSim**

MATSim is an open-source simulation toolkit implemented in Java. It has three features that make it a preferred platform over other travel demand modeling platforms. The first is the use of a **synthetic population that includes activity schedules** so that simulation incorporates activity scheduling behavior. The role of MATSim as a simulation of activity scheduling is discussed at great length in Chow (2018). A synthetic population provides much more detailed population details that can be sliced any number of ways (see He et al., 2020). The simulation captures the feedback loop between the traffic dynamics and user activity scheduling. The issue in many activity scheduling models is the lack of sensitivity to spatial temporal constraints reflected at a large scale in the population, a drawback discussed in Chow and Recker (2012) and Chow and Djavadian (2015). MATSim provides the feedback loop using a day-to-day adjustment process driven by a heuristic (a genetic algorithm).

The second is that MATSim can simulate traffic dynamics at a large scale using a **mesoscopic spatial queue model** (Cetin et al., 2003) for the traffic simulation.

Another advantage of MATSim is its ability to assess users' preferences for different modes and services relative to the congested traffic modes through a **day-to-day adjustment process**. Because MATSim is an open-source platform, there are many applications of MATSim around the world, including Berlin (Neumann, 2016; Ziemke, 2016), Zurich (Rieser-Schüssler et al., 2016), Singapore (Erath and Chakirov, 2016), among others. There is an online repository of existing open-source MATSim models around the world (MATSim, 2020). MATSim has also been used to evaluate several emerging technologies:
- Autonomous vehicle fleet (Hörl et al., 2019)
- Carshare (Ciari et al., 2016)
- Urban air mobility (Rothfield et al., 2018)
- Demand-responsive transit (Cich et al., 2017)
- Mobility as a Service (MaaS) (Becker et al., 2020)

As an agent-based simulation, MATSim can capture the behavior of each agent and the interaction between agents and transportation system. Each agent refers to an individual traveler. Traveler behavior is represented by a series of activities, travel modes and routes. MATSim uses an iterative framework to simulate the day-to-day process, as shown in Figure 2. The goal of the iterative framework is to find an equilibrated state of the system. The overall simulation procedures are:
- Put the agents with the initial travel plans into MATSim and simulate their mobility in the physical system.
- Calculate the score (utility) of each agent's executed plan.
- Randomly select a portion of agents and mutate their plans. Go back and re-run the simulation for the next "day" until the agents' scores converge.

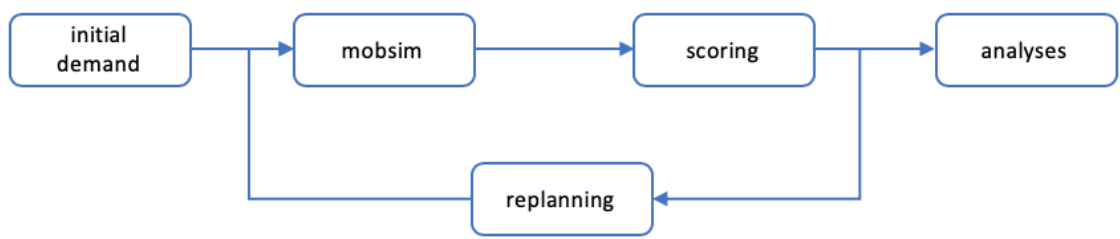

**Figure 2. Framework of simulation in MATSim (Horni et al., 2016).**

The output of MATSim contains the executed plan of all the agents. As such, many useful results can be extracted from it that cannot be captured by travel demand models using aggregate traffic assignment, such as:
- Individual mode shift for a specific scenario from a base scenario
- Departure time distributions across a day
- Trip travel distance distributions per mode
- Average hourly speed distribution across a day per link
- Transit ridership and load profile per route

- Passenger flow distribution per station by time of day
- Traffic count at a specific link by time of day

**2.3 Congestion pricing**

City-scale policies have large impacts on people's travel behaviors. Congestion pricing is a compelling example. Congestion pricing can benefit multiple stakeholders, like drivers, businesses, public transit, state and local governments, and society as a whole (Small, 1992; FHWA, 2019). In principle, congestion pricing done in a first-best basis (without any constraints) can shift travelers toward a system optimal state (Yang and Huang, 1998). In practice, it is not practically feasible to implement a first-best pricing scheme, and thus policymakers have resorted to second-best pricing schemes (see Chow and Regan, 2014, for a review).

According to the Federal Highway Administration (FHWA) of the US Department of Transportation, there are three main pricing strategies implemented or considered to implement in US: variably priced lanes, variable pricing on entire facilities, and cordon charges (Zhang and Yang, 2004). The I-15 interstate highway in San Diego is an implementation of variably priced lanes (Brownstone et al., 2003). Dynamic tolls are charged for vehicles using the High-Occupancy Toll lanes. Prices vary with the level of demand on the lanes. Lee County in Florida adopted the variable pricing on entire facilities (Burris and Swenson, 1998).

NYC proposed a plan for cordon charging (Baghestani et al., 2020). Former Mayor Bloomberg initially planned to charge a flat $8 daily price to passenger vehicles and $21 daily price to trucks from 6am to 6pm on weekdays (Schaller, 2010), although other plans have been proposed since. Under that plan, the charging zone is Manhattan under 86th street. This plan was projected to reduce vehicle miles traveled (VMT) in the charging zone by 6.7%. However, the plan was blocked in the State Legislature.

The NYC Department of Transportation (NYCDOT) has a congestion price plan in the area of Manhattan south of 60th street, as shown with the red border in Figure 3. The expected annual revenue is from $810M to $1.1B, which will be used to improve and maintain the public transit system in NYC (Holland and Shah, 2019).

Because the congestion pricing policy would vary by time of day, the travelers' elasticity to the price with respect to departure time should be modeled. Furthermore, the congestion pricing policy is evaluated in the context of ridership on other modes. However, best practice policy tools in NYC do not have sensitivity to departure time and do not consider emerging mobility services like FHVs and bikeshare.

## 3. Data

We use the MATSim platform to develop a simulation model for NYC, dubbed "MATSim-NYC". For any simulation model instance incorporating public transit in MATSim, the minimum inputs are a synthetic population file, a road network, the transit schedule, the transit vehicle parameters, and a configuration file of the global parameters.

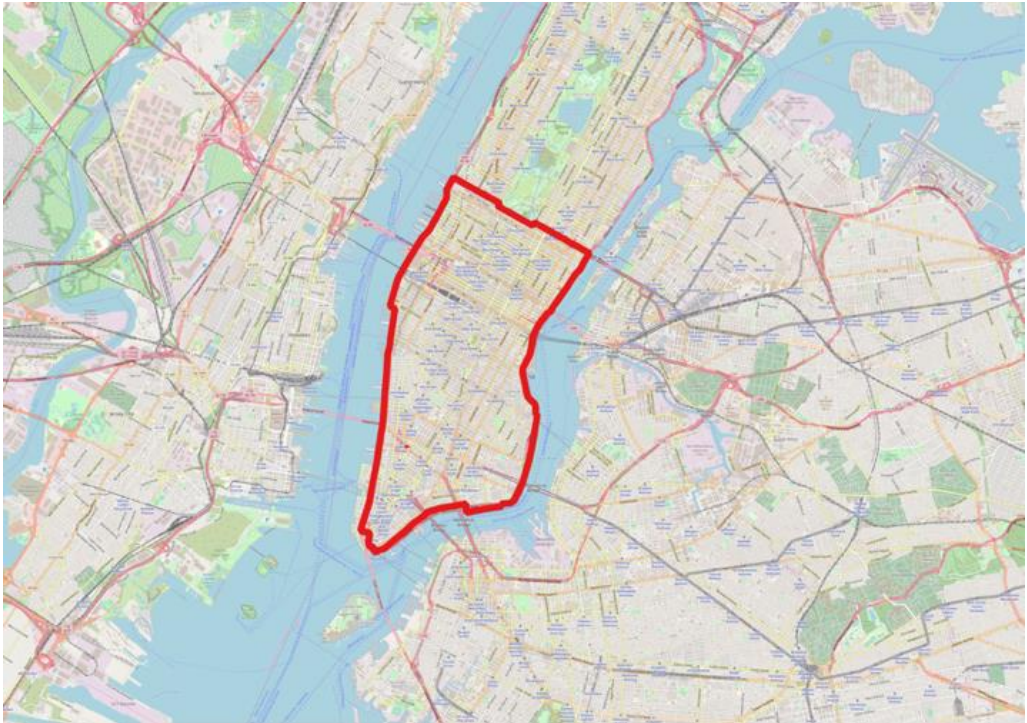
**Figure 3. Congestion pricing area in Manhattan (source: Open Street Map).**

**3.1 Synthetic Population**

We created an 8.24 million synthetic population for NYC, which incorporates people's daily travel agendas from the 2010/2011 Regional Household Travel Survey (RHTS) and mode choices at a subtour level calibrated to the base year 2016. A summary is provided with details of the synthetic population available in He et al. (2020).

*3.1.1 Zonal populations*

Based on the census data from the American Community Survey, 2040 Socio-Economic and Demographic Forecast data and 2016 Longitudinal Employer-Household Dynamics data, we generated a synthetic population of NYC in 2016 with PopGen (Ye et al., 2009). The synthetic population includes personal attributes like age, gender, school enrollment status, work status, and work industry, and household attributes like income group, household size, and number of cars owned. We extracted people's travel agendas from 2010/2011 RHTS, minus the mode of the trip made, and assigned them to individuals in the synthetic population according to their home locations and work or school enrollment status.

*3.1.2 Mode choice model*

A mode choice model was estimated to determine people's mode choice in subtour level according to the 2010/2011 RHTS and the trip count data of Citi Bike and For-Hire-Vehicle (FHV). The model has a nested structure and includes 8 modes: drive alone, carpool, public transit, taxi, bike, walk, FHV, and Citi Bike. The drive alone mode is determined at the tour level while the other modes are at the trip level. The model is initially estimated without FHV and Citi Bike.

We incorporated the emerging mobility services Citi Bike and FHV in the base mode choice model by duplicating the parameters (travel time and cost) from bike and taxi and perturbing the constant values to make the prediction of those trips as close to the observed trip counts as possible. Considering the correlation between those two modes and smartphone ownership, a smartphone ownership binary choice model was estimated and used as a feature in the Citi Bike utility function and as a determinant of FHV alternative availability. The estimated model was then used to simulate the modes of all the trips made in the synthetic population. The aggregated mode share of synthetic population was successfully validated against the 2017 Citywide Mobility Survey (CMS) as shown in He et al. (2020). The set of traffic analysis zones (TAZs) covering the synthetic population is shown in Figure 4.

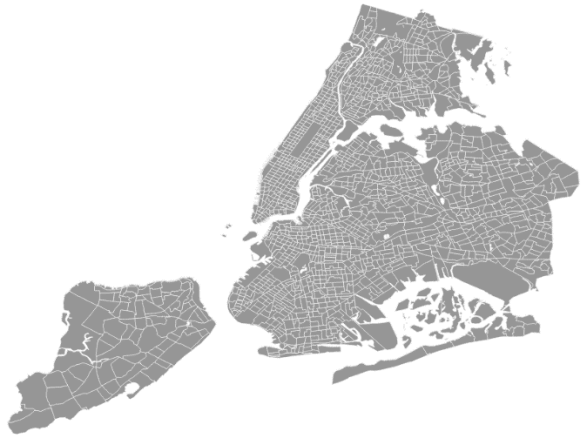

**Figure 4. TAZs for the synthetic population (source: NYMTC, 2000).**

*3.1.3 Synthetic trips of non-resident*

People who live outside the city but travel into the city for work need to be accounted for. A set of synthetic non-resident trips was duplicated from the 2010/2011 RHTS and expanded by the same population growth rate of 28.3% of the NYC residents from 2011 to 2016. The number of non-residents is 1,505,075, which makes up about 15.5% in the total simulation population. Furthermore, since our baseline model does not simulate trips outside the city, we identified gateway locations around the city and aggregated the trips to start from those gateways. The gateway locations are shown as red dots in Figure 5. These non-resident trips are assigned to the nearest gateway by mode, considering car, public transit, and walking.

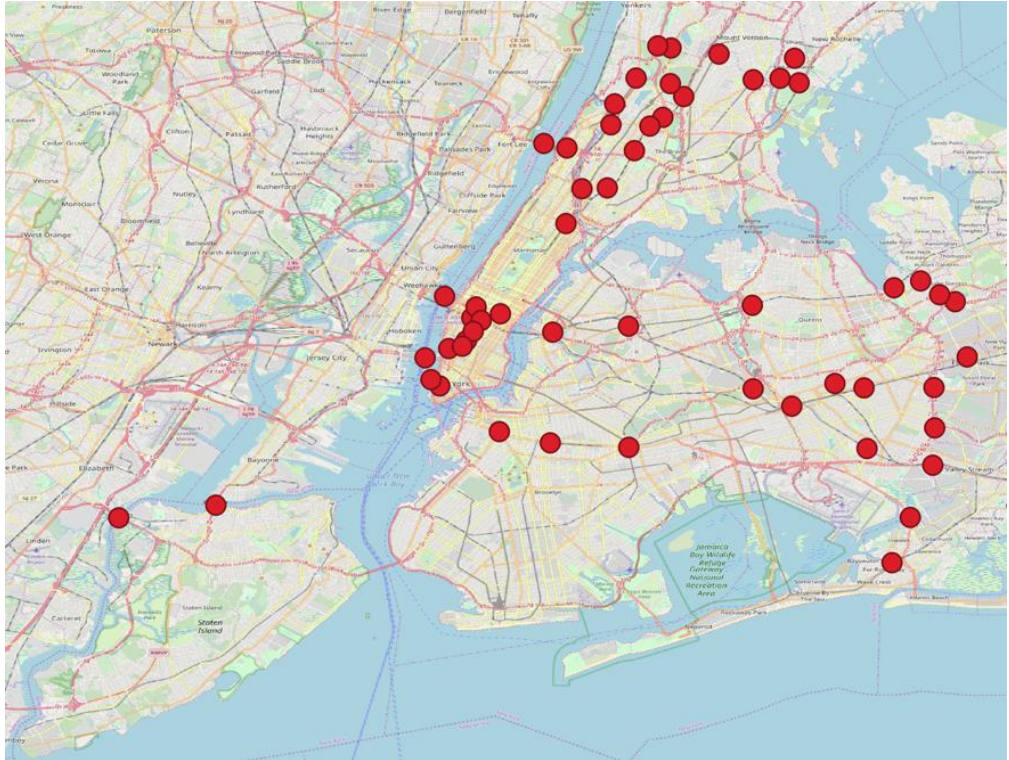
**Figure 5. Gateway locations around the city (source: Open Street Map).**

**3.2 Road Network**

The second input data for MATSim-NYC is the road network built from the OpenStreetMap (OSM) data. By using the open-source Java-based network edit tool JOSM (JOSM, 2018), we downloaded the OSM road network for NYC and transformed it to a MATSim formatted network shown in Figure 6. The link attributes included in the MATSim network are link length, capacity, free speed, number of lanes, and available modes. We kept the default link attributes the same with the OSM network. The links were classified into freeway link and arterial link by the free speed. If a link's free speed is greater than 33 m/s (around 74 mph), then the link is a freeway link; otherwise, it is an arterial link.

The network needs to be calibrated. We used 2016 bridges/tunnels volumes data (NYCDOT, 2016) and INRIX (https://inrix.com/) speed data in September 2016 from NYCDOT as references for calibration of the road network. The volume data consist of average hourly volumes across typical weekdays in 2016 for 59 bridges and tunnels in NYC while the speed data contain average observed speeds every five minutes for 6,996 link segments. We selected 18 bridges/tunnels around Manhattan as well as the Verrazano Narrows Bridge connecting Staten Island and Brooklyn as reference volumes for calibration. These locations are shown as the black bars in Figure 7.

We also defined a screenline along the East River as shown by the red curve in Figure 7. The screenline consists of Queensboro Bridge, Williamsburg Bridge, Queens Midtown Tunnel, Hugh Carey Tunnel, Manhattan Bridge, and Brooklyn Bridge. The Hudson River crossings are not considered since most of the trips are made by non-residents which are not sensitive to the calibration of the road network. The selected bridges and tunnels are listed in Table 1. For speed data, we took the average hourly speed on weekdays of September 2016 as reference. Figure 8 presents the distribution of average observed speed at 12 AM from that period which should reflect relatively free flow speeds.

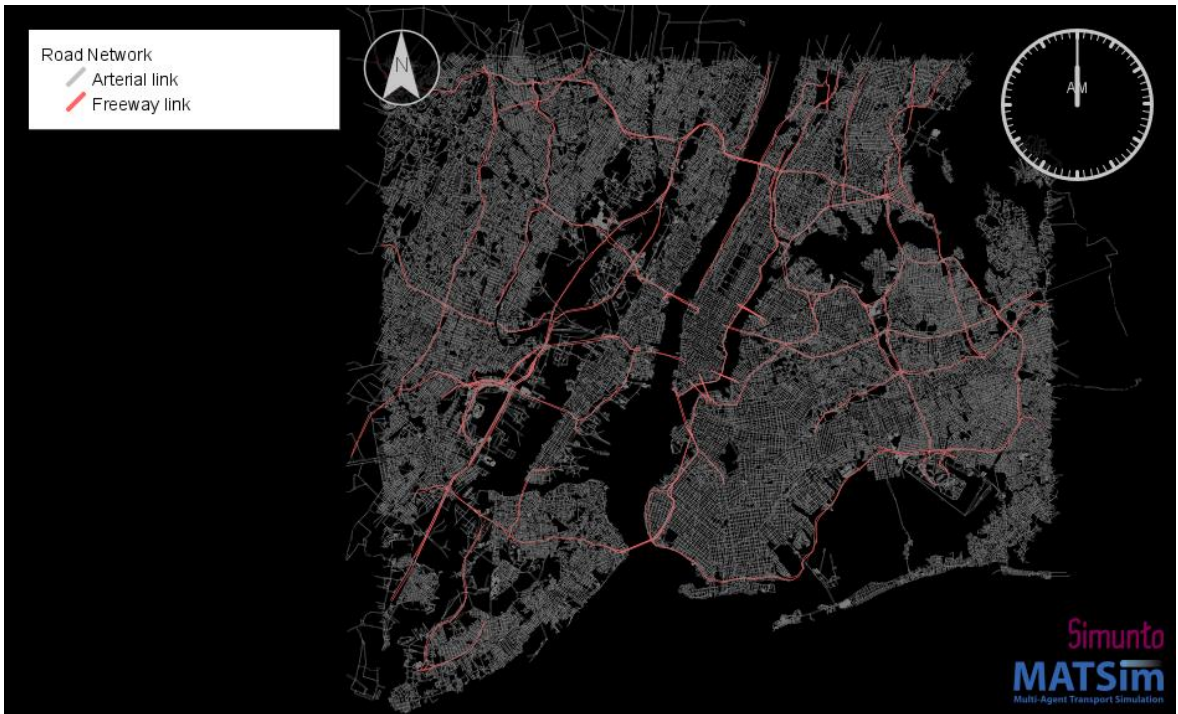
**Figure 6. OSM road network for MATSim-NYC.**

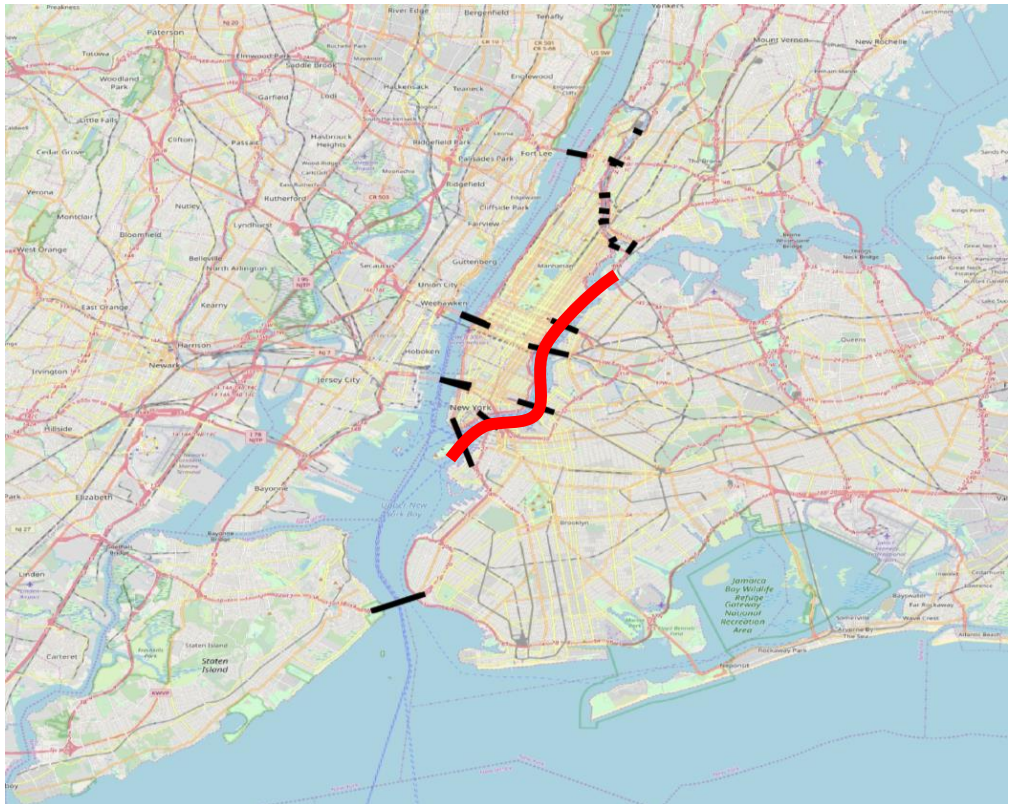
**Figure 7. Locations of Bridge/Tunnel selected (source: Open Street Map).**

**Table 1. List of traffic count facilities for network calibration**

| ID | Facility | ID | Facility | ID | Facility |
|---|---|---|---|---|---|
| 1 | Brooklyn Bridge | 8 | Washington Bridge | 15 | Verrazzano-Narrows Bridge |
| 2 | Ed Koch Queensboro Bridge | 9 | Willis Avenue Bridge | 16 | George Washington Bridge |
| 3 | Williamsburg Bridge | 10 | 145th Street Bridge | 17 | Holland Tunnel |
| 4 | Alexander Hamilton Bridge | 11 | Hugh L. Carey Tunnel | 18 | Lincoln Tunnel |
| 5 | Macombs Dam Bridge | 12 | Queens-Midtown Tunnel | 19 | Manhattan Bridge |
| 6 | Madison Avenue Bridge | 13 | Robert F. Kennedy Memorial Bridge Manhattan Plaza | | |
| 7 | University Heights Bridge | 14 | Robert F. Kennedy Memorial Bridge Bronx Plaza | | |

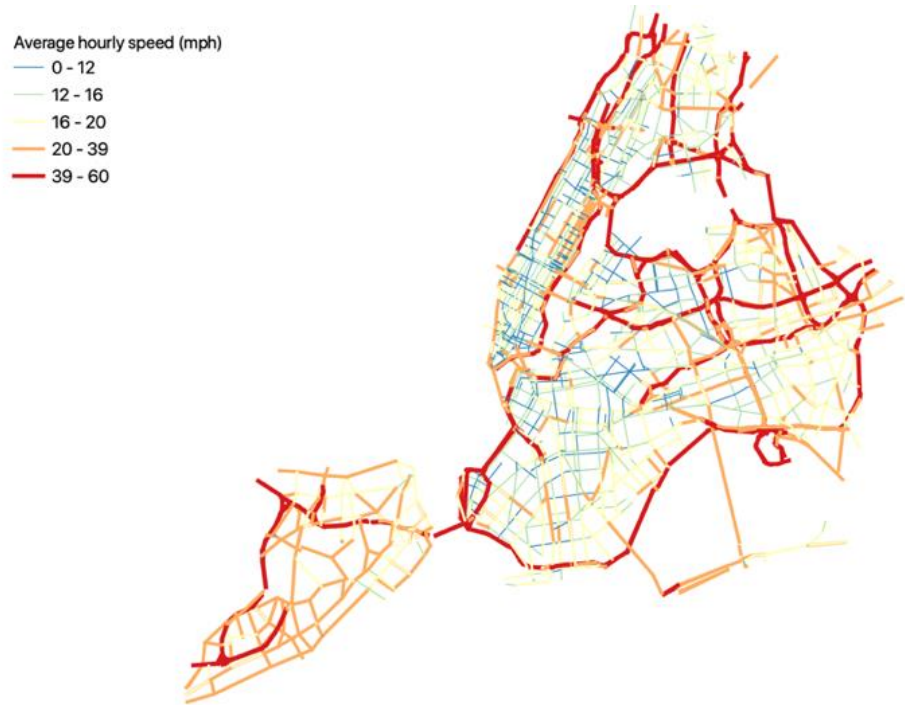

**Figure 8. Distribution of average hourly speed at 12 AM in September 2016 from INRIX data.**

### 3.3 Public Transit Network and Schedule

The public transit network is generated from the General Transit Feed Specification (GTFS) data (MTA, 2018). We selected the historical GTFS data in September 2016 and mapped it with the road network. While the MATSim platform can model shared right-of-way transit (e.g. buses) to use the same links as cars, experience shows it would significantly increase computation time. As a result, we treated transit and car use as separate links, assuming that the schedules are adequately up to date to reflect the recurrent congestion.

The transit schedule consists of stop locations, routes, and corresponding timetables for each transit line. We combined the road network and transit network into a modal network with the class "PublicTransitMapper" in pt2matsim extension of MATSim, as shown in Figure 9. Vehicle capacities (for the fourth input data file) are determined by vehicle types (ERA, 2016) and MTA's services standards (MTA, 2018a).

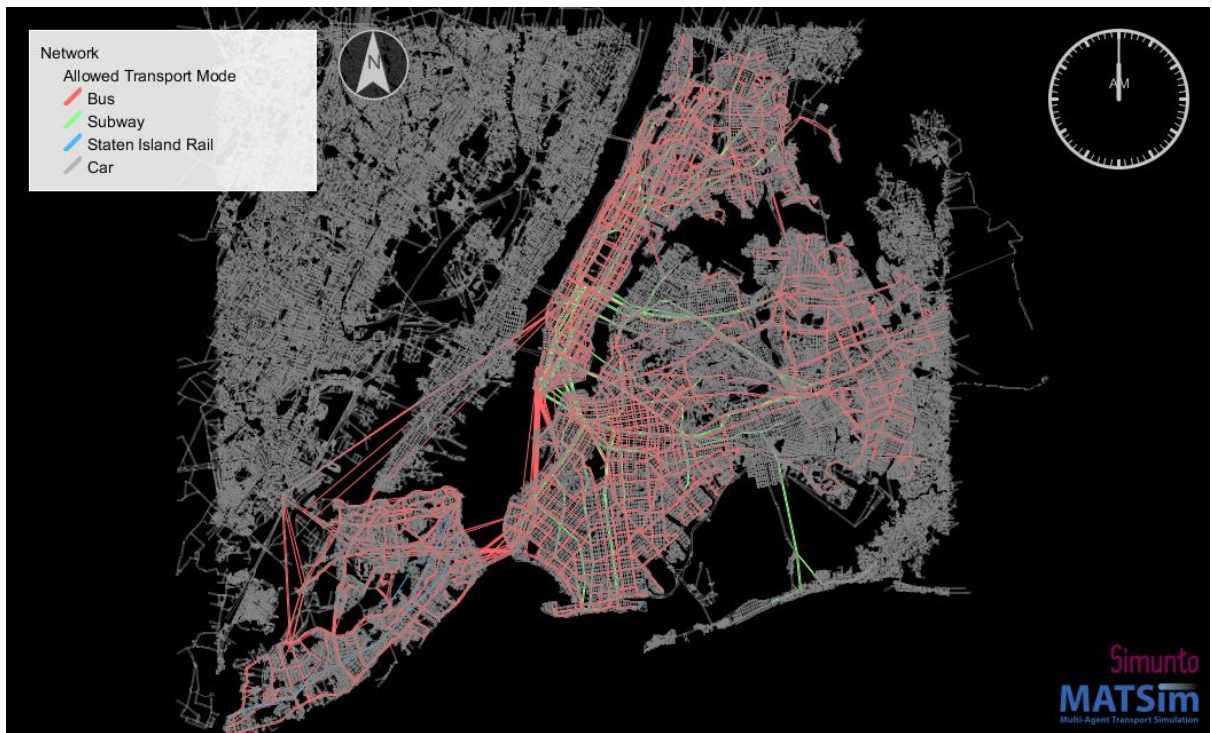
**Figure 9. Modal network for MATSim-NYC.**

### 3.4. Data for validation of MATSim-NYC

In summary, MATSim-NYC is defined by the 2016 synthetic population, OSM network calibrated from crossings and INRIX speed data, and GTFS data from 2016. It includes 8 modes of travel, where only car and transit operate on a modal network where traffic dynamics can cause congestion. This implies that MATSim-NYC's other modes do not feature any supply-side congestion, although the congestion on the road would introduce cross-elasticities in the demand.

Two datasets were used to validate the base MATSim-NYC simulation model. The first is the 2016 Average Weekday Subway Ridership data (MTA, 2018b). This dataset provides the average weekday ridership per station around the city. The ridership is defined as the number of passengers entering the station (MTA, 2018c). Ten stations were selected for validation and the locations are shown in Figure 10. These ten stations were chosen because they have relatively high daily ridership in Manhattan, Brooklyn and Queens: they represent 19.3% of the total ridership in the city.

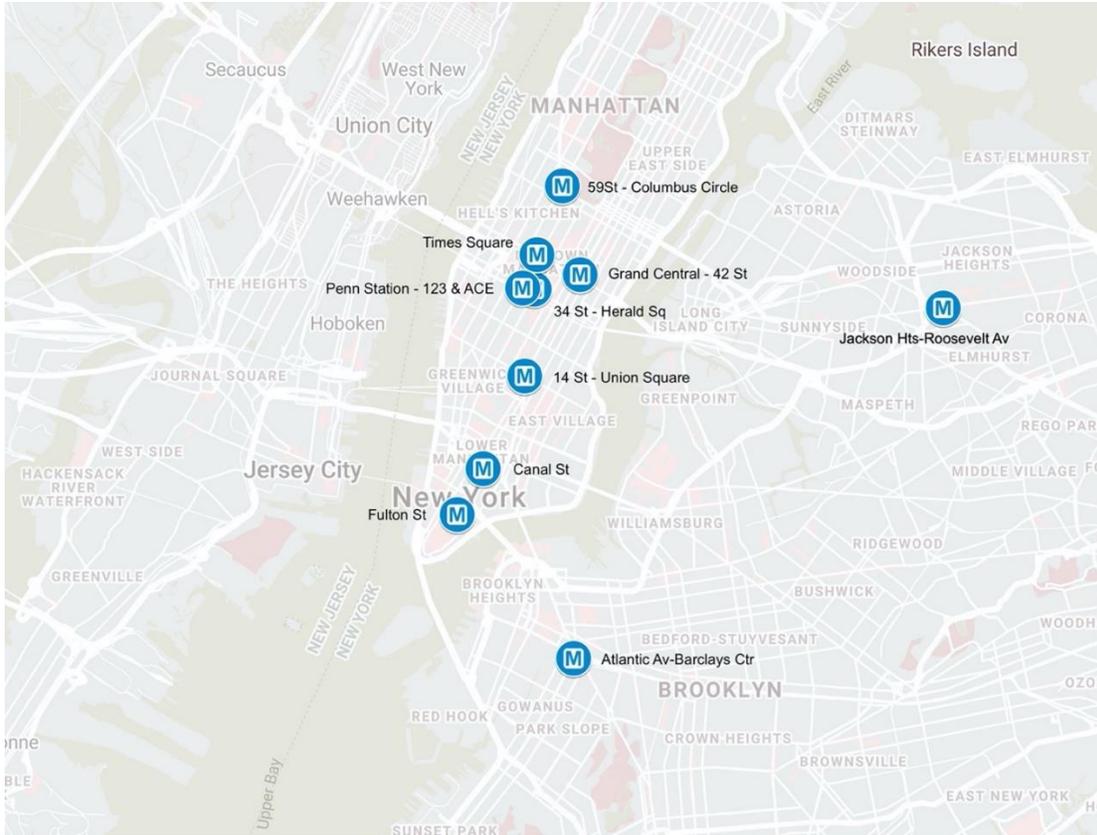
**Figure 10. Selected subway stations for validation (source: Google Maps).**

The other dataset is the 2014-2018 Traffic Volume Counts data (DOT). This dataset was collected by NYC DOT for validation of the NYBPM, chosen to have a consistent validation effort with their model. Hourly volumes of 597 locations from 2014 to 2018 around the city were provided. Volumes of 46 locations in 2016 were selected for validation. These were chosen because the locations are the main freeways or arterial roads in Manhattan, Brooklyn and Queens among all the locations in the dataset. We matched those locations to links in the MATSim-NYC network shown in Figure 11.

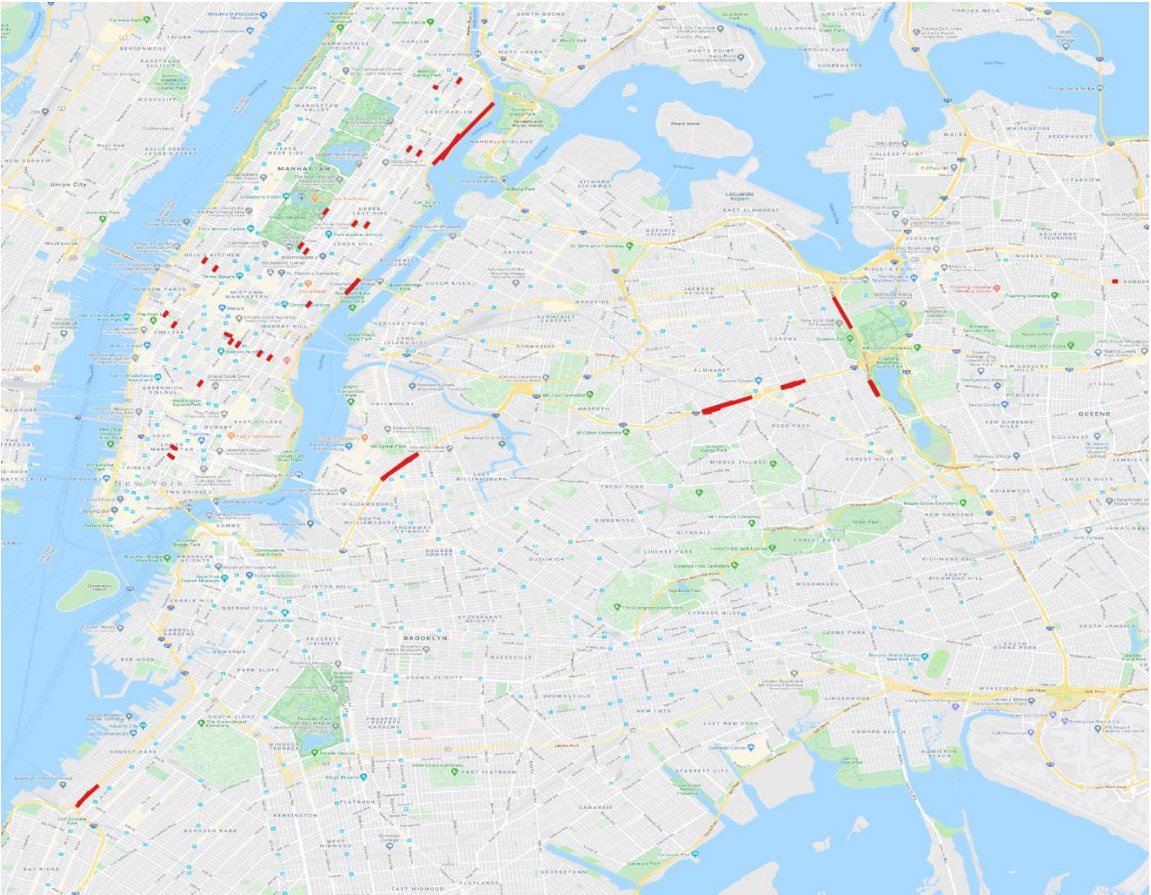
**Figure 11. Selected links for volume validation (source: Open Street Map).**

## 4. Calibration and validation of MATSim-NYC

Several of the input data need to be calibrated for the output of MATSim-NYC to best match observed data introduced earlier. Due to computational cost, a scaled sample of the population is typically used for the simulation. For example, Berlin (Ziemak et al., 2019), Zurich (Balmer et al., 2008), and Paris (Hörl et al., 2019) use 10% scaled populations in their simulations. Because of the size of NYC, we use a 4% sample population (~320K agents, selected based on home TAZ) in the simulation and calibrate the road network flow capacity and transit vehicle capacities accordingly. The translation is not linear, i.e. it is not expected that transit capacities scaled to 4% would fit the 4% population.

For the road network, there is both the flow capacity and the unsaturated speed that need to be calibrated. The preferences of the population agents also depend on fitted utility functions. Since MATSim uses a single-level multinomial logit (MNL) structure of evaluating different mode alternatives, there needs to be a conversion of the nested logit model parameters to an equivalent MNL model.

## 4.1. Configuration parameters in MATSim and other features

### 4.1.1. Travel parameters

Since the default MATSim only supports mode choice with the Multinomial Logit (MNL) model, we adjusted the tour-based nested logit model (see He et al., 2020) into an equivalent trip-based MNL model as shown in Eqs. (1) – (3). The smartphone ownership model and the availability of Citi Bike attribute from He et al. (2020) are ignored in MATSim.

$$V_{driving,k} = \beta_{driving,C}\mu + \beta_{cost}Cost_{driving,k} \quad (1)$$

$$V_{j,k} = \beta_{j,C} + \beta_{cost}Cost_{j,k} + \beta_{j,time}Time_{j,k}, \quad j \in J\setminus\{0,1\} \quad (2)$$

$$V_{2,k} = \beta_{2,C} + \beta_{cost}Cost_{2,k} + \beta_{2,time}(Time_{2,k} - Time_{driving,k}) + \beta_{AT}AT_k + \beta_{ET}ET_k + \beta_{WT}WT_k \quad (3)$$

where $\mu$ is the scale parameter from the nested logit model, $Cost_{driving,k}$ is the cost of driving, $Time_{j,k}$ is the travel time, $AT_k$ is the access time, $ET_k$ is the egress time, and $WT_k$ is the transfer time, all for trip $k$. Modes are numbered as $J = \{$0: drive alone, 1: carpool, 2: transit, 3: taxi, 4: bike, 5: walk, 6: FHV, 7: Citi Bike$\}$.

The equivalent parameters used for the travel score in MATSim-NYC are shown in Table 2 for two population segments: Manhattan and non-Manhattan. The segments refer to residents who live in Manhattan versus those who live in one of the boroughs outside Manhattan. The parameters are explained in He et al. (2020). Some of the parameters are set to zero because they were not statistically significant. As recommended in the user guide for MATSim (Nagel et al., 2016), the travel time parameter of car was normalized to zero and the travel time parameters of the rest of the modes were adjusted accordingly by subtracting the travel time parameter of car (including the parameters of Access Time, Egress Time, and Transfer Time). All the travel times (including the Access Time, Egress Time and Transfer Time) are transformed into hours. The transit fare is set to $2.75 per trip.

**Table 2. Parameters for travel score of both Manhattan and Non-Manhattan segments**

| | Manhattan | car | carpool | transit | taxi | bike | walk | FHV | Citi Bike |
|---|---|---|---|---|---|---|---|---|---|
| | Constant | -0.06 | 0.00 | 2.95 | 1.06 | 0.44 | 5.73 | 0.79 | -0.37 |
| | Travel Time | 0 | 2.35 | 0.00 | 1.75 | -2.55 | -3.94 | 1.75 | -2.55 |
| | Cost | -0.06 | | | | | | | |
| transit | Access Time | -0.96 | | | | | | | |
| transit | Egress Time | -0.86 | | | | | | | |
| transit | Transfer Time | -1.46 | | | | | | | |
| | Non-Manhattan | car | carpool | transit | taxi | bike | walk | FHV | Citi Bike |
| | Constant | -0.05 | 0.00 | 0.76 | -1.81 | -1.35 | 3.49 | -3.38 | -2.04 |
| | time | 0.00 | 0.36 | 0.00 | 0.00 | -5.64 | -5.05 | 0.00 | -5.64 |
| | cost | 0 | | | | | | | |
| transit | Access Time | -1.71 | | | | | | | |
| transit | Egress Time | -1.67 | | | | | | | |

| | Transfer Time | -1.61 |
|---|---|---|

As noted in He et al. (2020), the nested logit model parameters led to an estimated value of time (VOT) of $29/h for Manhattan residents. We use this value to estimate the consumer surplus impacts of the congestion pricing policy for all of NYC, noting that this value likely is an upper bound.

### 4.1.2. Activity parameters

In MATSim, each agent has multiple plans in a day and selects one to execute among them. Several selection strategies can be adopted (see Horni and Nagel, 2016). In our simulation, we adopted the "SelectExpBeta" strategy to perform Multinomial Logit Model selection between plans according to Eq. (4). $C_n$ is the set of plans for agent $n$, $S_{ni}$ is the score of plan $i$ for agent $n$, and $\mu$ is the controlling parameter for the preference of higher scores, which is set to 1. While the algorithm uses an MNL structure to select a plan, the plan set is an outcome of the algorithm and not representative of a real choice set available to a traveler, so any logsum analysis of the plan set itself would not have any economic interpretation.

$$P_n(i|C_n) = \frac{e^{\mu S_{ni}}}{\sum_{j \in C_n} e^{\mu S_{nj}}} \tag{4}$$

The score of a plan is similar to the mode utility in the mode choice model but incorporates the additional utility (score) of activities (Nagel et al., 2016). The basic function of calculating the plan score is shown in Eq. (5).

$$S_{plan} = \sum_{q=0}^{N-1} S_{act,q} + \sum_{q=0}^{N-1} S_{trav,mode(q)} \tag{5}$$

where $N$ is the number of activities in the plan, $S_{act,q}$ refers to the score of activity $q$ and $S_{trav,mode(q)}$ represents the score of trip after activity $q$ via $mode(q)$. The last activity is combined with the first one to have the same number of activities and trips. The activity score is broken down into other components as shown in Eq. (6), where the duration score $S_{dur,q}$ and late arrival score $S_{late\ arr,q}$ are computed using Eq. (7) – (8).

$$S_{act,q} = S_{dur,q} + S_{late\ arr,q} \tag{6}$$

$$S_{dur,q} = \beta_{dur} t_{typ,q} \ln(t_{dur,q}/t_{0,q}) \tag{7}$$

$$S_{late\ arr,q} = \begin{cases} \beta_{late\ arr}(t_{start,q} - t_{late\ arr,q}), & \text{if } t_{start,q} > t_{late\ arr,q} \\ 0, & \text{otherwise} \end{cases} \tag{8}$$

where $t_{typ,q}$ (in hours) is the typical duration of activity $q$, $t_{dur,q}$ is the actual duration of activity $q$, $t_{0,q}$ is the duration when the utility of activity $q$ starts to be positive. $t_{0,q}$ is set to $t_{typ,q} \exp(-10/t_{typ,q})$ according to Rieser et al. (2014). $t_{start,q}$ is the actual start time of activity $q$, $t_{latest\ arr,q}$ is the latest start time of activity $q$ without penalty. Eq. (7) defines the score of performing an activity, which is usually positive. An activity may have a negative score if its duration is less than $t_{0,q}$. Eq. (8) is the penalty of late arrival.

The activity diary data from the NYMTC Household Travel Survey only includes revealed travel without planned travel preferences and for only one day per respondent. Without planned trip data to compare to the revealed trip data or panel data across multiple days (e.g. as done in Axhausen et al., 2002), the parameters for early/late arrival penalties and activity durations cannot be properly estimated. As such, we make assumptions to ensure that the late arrival penalties and the activity duration penalties are higher than travel time so that the maximum activity durations and non-late arrivals are always preferred under congestion. Early arrival penalty is assumed to be zero, which is why it is omitted in Eq. (6). This calibration ensures that the model is sensitive to trade-offs between traffic congestion and schedule delays. However, the $S_{act,q}$ portion of the utility scores would not have any economic interpretation since they are not fitted to the data and need to be left out of the consumer surplus comparisons in Section 5.2 Results.

We have five activity types defined in MATSim: Home, Work, School, University, and Secondary. We set the maximum durations of the activity types as 8h, 8h, 8h, 1h and 1h, respectively. $\beta_{dur}$ is set to be the same with the parameter for travel time of driving alone but positive, and the $\beta_{late\ arr}$ is determined relative to the parameter of travel time of driving alone according to Small (1982).

Small (1982) proposed a scheduling model for work trips and the results indicated that the parameter of schedule delay is about 2.39 times of travel time. We use this to estimate a $\beta_{late\ arr} = -4.19$. Since $\beta_{dur}$ is assumed in order to have agents try to maximize their durations, we make sure to omit the duration attribute from the score calculations when determining consumer surplus.

### *4.1.3. Other considerations*

We also made the following changes to further calibrate MATSim-NYC:
1) We added parking cost for car trips in the simulation. The average values are set to $5.19 per trip. This cost comes from the average parking cost collected from 2010/2011 RHTS.
2) We charged tolls for each vehicle that traversed a toll link. The tolls charged are the same as the Metropolitan Transportation Authority (MTA) and the Port Authority of New York and New Jersey (PANYNJ). PANYNJ facility tolls (Lincoln Tunnel, Holland Tunnel, George Washington Bridge, Bayonne Bridge, Goethals Bridge, Outerbridge Crossing) are set to $12.50 during peak hours (6-10AM, 4-8PM) and $10.50 in off-peak hours, collected in the inbound direction toward NYC. MTA facilities (Bronx-Whitestone Bridge, Throgs Neck Bridge, and Robert F. Kennedy Bridge; Hugh L. Carey Tunnel and Queens Midtown Tunnel) are charged $6.12 in both directions.
3) For the non-residents of the city, we didn't apply mutation strategies like "subtour mode choice", "reroute" and "time mutate" to the non-resident subpopulation. They are treated as background traffic in our simulation.
4) Since a 4% scaled population is used, we adjust the flow capacity accordingly. Different flow capacity factors (0.04, 0.05, 0.10 and 0.15) were experimented with. A value of 0.15 was needed to ensure total simulated volumes of selected count facilities were close to the real traffic count. Therefore, in the simulation, the 4% sample population and 0.15 flow capacity factor were adopted.
5) Some modes were simulated in the multimodal network while others were "teleported", due to the computation time concerns. Teleportation means that the performance level of that mode is not impacted by traffic dynamics directly, though they do interact through the day-to-day responses. The car, taxi, FHV and transit modes were simulated in the network. Other modes including bike, walk, carpool and Citi Bike were teleported outside the network.
6) A sensitivity test was conducted to ensure the stability of a single simulation using the 4% population. Four different runs of the baseline scenario were conducted and the standard deviation of predicted trips per mode across the four runs was at most 3.6% of the mean. This suggests that the 4% population was sufficiently stable.

All the data, code, and files are shared on Zenodo (He, 2020).

### 4.2. Network calibration

#### *4.2.1. Calibration framework*

MATSim adopts a queue-based approach to simulate traffic flow (Horni et al., 2016; Dobler and Axhausen, 2011). In the queue-based model, the flow capacity, storage capacity, and free flow speed link travel time are taken into consideration (Agarwal et al., 2015). The queue-based model regards a network link (i.e. a road segment) as a point queue. When a vehicle enters a link, it is added to the tail of a waiting queue at the start of the link. The travel time on that link is set based on a constant unsaturated flow speed. The resulting traffic fundamental diagram of the queue-based model is presented in Figure 12.

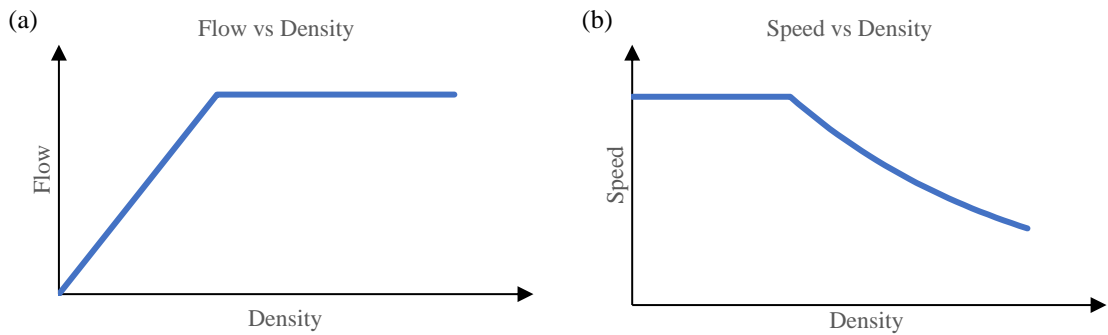

**Figure 12. MATSim fundamental diagram: (a) Flow vs Density and (b) Speed vs Density.**

The queue-based model ignores the intra-link interactions to improve the computational efficiency. When a link is saturated, the flow stays the same in MATSim as long as the downstream link is open, whereas the flow in the real world goes down as density increases. This assumes that the MATSim equilibrium operates in a steady state where traffic would not be oversaturated. Other factors (e.g. arterial traffic system, non-resident (tourist) trips, truck deliveries, etc.) that affect the link flow capacity and free flow speed are not incorporated due to limitations of the data. The road network attributes (link flow capacity and free flow speed) need to be calibrated.

We define two sets of parameters for those major network attributes. The unsaturated flow speeds are first calibrated according to INRIX speed data. The link capacity factors are then perturbated iteratively to a closer volume distribution compared to the real traffic count. The overview of the process is shown in Figure 13.

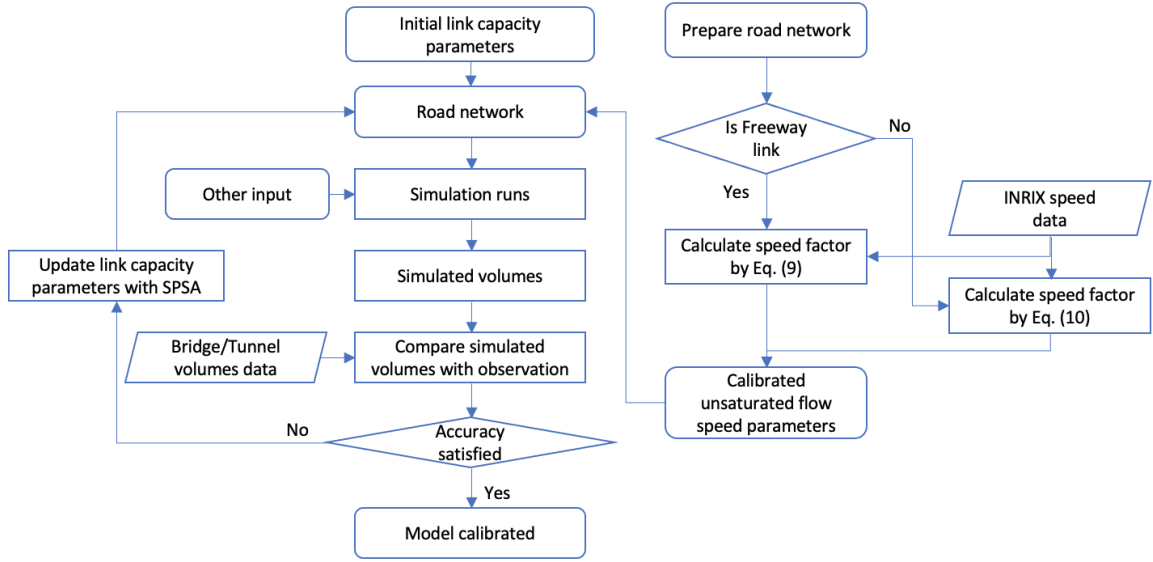

**Figure 13. The calibration process.**

### 4.2.2. Unsaturated flow speed calibration

The speed from INRIX data varies with time. Unsaturated flow speeds are calibrated for six time periods: 6 - 9 AM, 9 AM - 12 PM, 12 - 3 PM, 3 - 6 PM, 6 - 9PM and 9 PM - 6 AM. Different speed factors are applied to links according to link types and time. Let $L = \{1,2\}$, where 1 is freeway and 2 is arterial, and $T = \{1,2,3,4,5,6\}$ refers to the six time periods correspondingly.

First, the average speed data from INRIX were aggregated by link type and time period (Table 3). We can calculate the link speed factors of each freeway link as Eq. (9).

$$f^s_{1,t} v^0_1 = v^{ob}_{1,t}, t \in T \tag{9}$$

where $v^{ob}_{1,t}$ refers to the average observed speed for the freeway link in time period $t$, and $v^0_1$ represents the default unsaturated flow speed for the freeway link.

**Table 3. Average observed speed in each time period for Freeway and Arterial links (mph)**

|  | 6AM-9AM | 9AM-12PM | 12PM-3PM | 3PM-6PM | 6PM-9PM | 9PM-12AM |
|---|---|---|---|---|---|---|
| Freeway | 36.88 | 37.93 | 37.61 | 33.05 | 36.25 | 42.41 |
| Arterial | 14.10 | 13.42 | 13.11 | 12.80 | 13.91 | 15.34 |

For Arterial links, further adjustments are needed due to high variation in free flow speed. A set of sub-categories $J = \{1,2,3\}$ is used to represent arterial links in OSM with base unsaturated flow speeds 22.2 m/s, 15.0 m/s and 8.3m/s, respectively. To adjust the link speed as close to the observed speed as possible, we apply different link speed factors to corresponding sub-categories of arterial links using Eq. (10) to match to INRIX data. The final average arterial speeds are shown in Table 3.

$$f^s_{2,t,j} v^0_{2,j} = v^{ob}_{2,t}, t \in T, j \in J \tag{10}$$

where $v^0_{2,j}$ is the default link free speed for link sub-category $j$.

### 4.2.3. Link capacity calibration

After the unsaturated flow speeds are adjusted to accommodate time-varying conditions like non-resident traffic and signal control, the next step is the link capacity calibration. Based on the time variant network feature of MATSim, a link behaves according to Eq. (11).

$$C^{sim}_{l,t} = C^0_l f^c_{l,t} \tag{11}$$

where $C^{sim}_{l,t}$ is the link flow capacity in simulation for facility type $l \in L$ in time period $t \in T$, $C^0_l$ is the default link capacity for facility type $l \in L$ from OSM, and $f^c_{l,t}$ is a factor to adjust the capacity for a given facility type $l \in L$ in time period $t \in T$. The storage capacities are unchanged from the default values (average vehicle length set to 7.5 m).

The flow capacities are used in a cellular automata model to propagate traffic within the road network to output location volumes by time of day. The simulation then proceeds through multiple days with each subsequent day updating the travel choices of the population and the cellular automata model updates the system performance based on the propagation of the new day's traffic. The resulting output of $n$ days of simulation is denoted as $\Omega_n$, i.e. $(\{V^{sim}_{i,t}\}) = \Omega_n(\theta, I; f^s)$ for location $i$ at time period $t$. In our model there are 12 (2 types, 6 periods) capacity parameters to be calibrated. The 19 locations in Table 1 are used for the calibration.

The SPSA method (Spall, 1988, 1998a, 1998b) is used to calibrate the capacity parameters. The SPSA method is shown in Algorithm 1, where $\hat{\theta}_k$ generically represents the estimated link capacity factors in the $k^{th}$ iteration of calibration.

**Algorithm 1: SPSA (source: Spall (1988, 1998a, 1998b))**

| Input: initial vector of link capacity factors $\hat{\theta}_0$ |
|---|
| 0: Initialization |
|     Set $k = 1, \beta, \gamma, a, A, c$ as initial values. |
| 1: Generation of simultaneous perturbation vectors |
|     Generate a $p$-dimension $\Delta_k$ by Monte Carlo. Each element of $\Delta_k$ is generated independently from a Bernoulli $\pm 1$ distribution with probability of ½. |
|     Calculate $a_k = \frac{a}{(k+A)^\beta}$ and $c_k = \frac{c}{k^\gamma}$. |
| 2: Gradient approximation |
| 3: Update parameter $\hat{\theta}_k$ estimation |
| Terminate the algorithm when loss function reaches a threshold or the maximum iteration is reached, else set $k = k + 1$ and return step 1. |
| Output: final vector of link capacity factors $\hat{\theta}_k$ |

The SPSA method is implemented in Java and run with the MATSim-NYC model. The initial coefficients are $\beta = 0.602, \gamma = 0.101, a = 0.16, A = 3000, c = 0.05$. The initial value of the link flow capacity factor is set to $\hat{\theta}_0 = 0.6$. The upper bound and lower bound of link capacity factors are set as 0.8 and 0.3 to make sure the link capacities stay in a reasonable range. To save computation time for calibration, we run only $n = 50$ days in MATSim for each iteration of the calibration.

## 4.3. Calibration Results

The calibration was run on a desktop with Intel(R) Xeon(R) CPU E5-2637 v4 @ 3.50GHz processors and 128GB RAM for 6 iterations until the simulated screenline volumes were observed to be within 5% of the observed data. Each iteration took around 11 hours. The relative errors between observed and simulated screenline volumes per iteration are presented in **Figure *14*** and the calibrated link capacity factors are presented in Table 4.

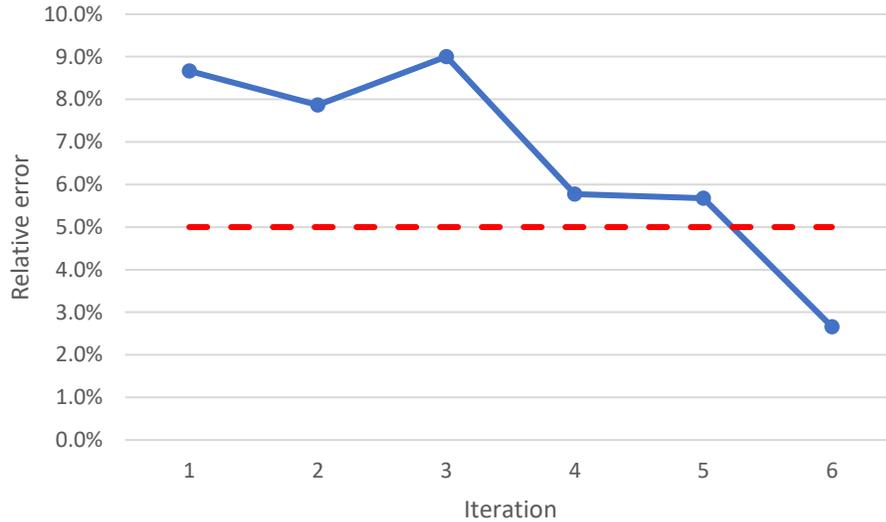

**Figure 14. Relative error between observed and simulated screenline volumes per iteration.**

**Table 4. Calibrated link capacity factors for Freeway and Arterial links**

|  | 6AM-9AM | 9AM-12PM | 12PM-3PM | 3PM-6PM | 6PM-9PM | 9PM-12AM |
|---|---|---|---|---|---|---|
| Freeway | 0.61 | 0.69 | 0.51 | 0.63 | 0.55 | 0.68 |
| Arterial | 0.59 | 0.51 | 0.57 | 0.50 | 0.46 | 0.68 |

As Figure 15 shows, the simulated average speed in each time period is close to the INRIX speed. For the Freeway links, the relative difference is 7.2% on average, with the highest at 9.5%. For the Arterial links, the relative difference is 17.1% on average, with the highest at 18.5%. The results of the Arterial links are worse because the default free flow speeds of the Arterial links vary more.

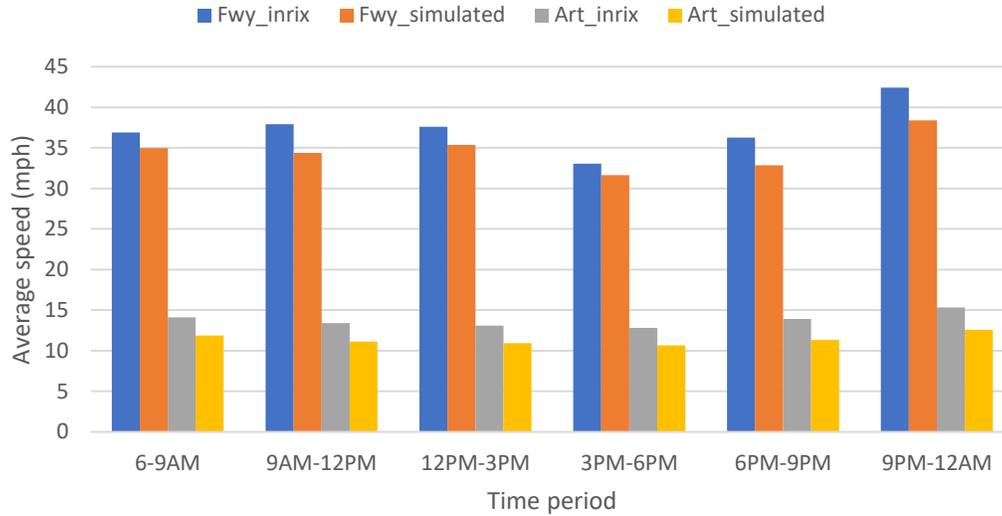

**Figure 15. Comparison of INRIX speed (observed) and simulated speed.**

The simulated volumes of the East River screenline are also close to the real traffic counts. The average difference between the total daily simulated volumes and real volumes is only $+1.8\%$, as shown in Figure 16. If we look at different time periods, the relative difference is 10.3% on average, with the highest at 17.6%. We also compare the calibration results of the screenline with the NYBPM 2010 update (Brinckerhoff, 2014). The MATSim NYC model has a difference of $+1.8\%$ from the observed data compared to a $-2.4\%$ with the NYBPM.

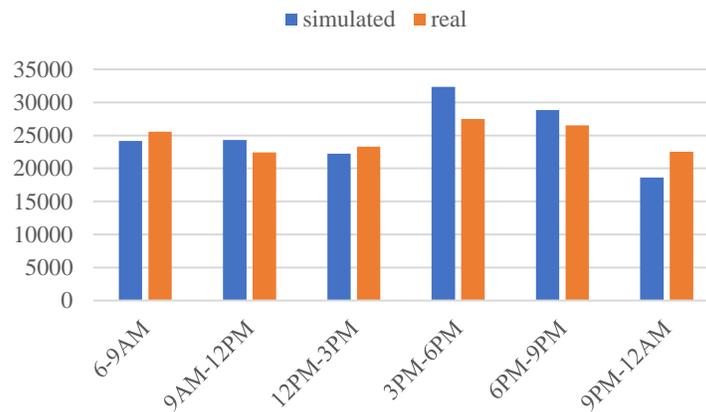

**Figure 16. Average simulated and real volume distribution of the East River screenline across all time periods.**

### 4.4. MATSim-NYC validation

We validated the calibrated model by comparing the simulation output subway station ridership with observed turnstile count data as well as with volumes on key road corridors. The comparison of subway station ridership is presented in Table 5. The average difference between simulated and observed daily ridership per station is 8%. Most of the stations have close predictions to the observations, except for Penn Station, Atlantic Av-Barclays Ctr in Brooklyn, and Jackson Hts-Roosevelt Av Queens. The high deviations in ridership of Penn Station, Atlantic Av-Barclays Ctr might be due to passengers arriving at railway stations nearby. Our base model overestimated passengers coming to the city by rail, which led to the higher simulated ridership. The Jackson Hts-Roosevelt Av station has 21% lower prediction, but the absolute difference is not very large.

**Table 5. Comparison of daily ridership of subway stations**

| Station | Real | Simulated | Difference % |
| --- | --- | --- | --- |
| 14 St - Union Square | 106,718 | 97,825 | -8% |
| Grand Central - 42 St | 158,580 | 170,025 | 7% |
| Penn Station - 123 & ACE | 173,108 | 256,825 | 48% |
| Times Square | 202,363 | 191,425 | -5% |
| Fulton St | 85,440 | 83,025 | -3% |
| Canal St | 70,806 | 78,250 | 11% |
| 59St - Columbus Circle | 73,836 | 75,050 | 2% |
| 34 St - Herald Sq | 125,682 | 124,500 | -1% |
| Atlantic Av-Barclays Ctr (Brooklyn) | 42,711 | 59,350 | 39% |
| Jackson Hts-Roosevelt Av (Queens) | 52,296 | 41,200 | -21% |
| **Average** | **1,091,540** | **1,177,475** | **8%** |

Among the 42 links in the MATSim-NYC road network selected to validate the volumes, the average relative difference between simulated volumes and real counts is 39.8%. The median difference is 29%, which suggests a few large outliers but otherwise the model is relatively well-behaved for a citywide simulation across multiple modes to be used for quick response analysis of emerging technologies and policies. For example, Flyvbjerg et al. (2005) note that average transportation planning model forecasts have errors on the order of 20%+, and that is for local transportation projects as opposed to citywide multimodal models.

# 5. Congestion pricing case study

## 5.1. Scenario parameters

We study congestion pricing using MATSim-NYC. The advantage of using MATSim-NYC to evaluate these scenarios is that travelers can experience queue delays in a single day's simulation due to changes in spillback effects from the congestion pricing and can then adjust their mode and departure time choices in subsequent days in the simulation. This ensures both spatially heterogeneous modal and temporal elasticities compared to a static model which forces travelers to, at best, substitute trips with other modes or routes in the same time period. As the model is multi-agent and linked up with the experienced travel agendas on the dynamic traffic network, we can further track the mode shifts individually and quantify the output utility change to each traveler to determine social welfare effects of a policy on different population segments. This has powerful economic value to evaluating emerging transportation technologies and policies.

The RPA published a report (RPA, 2019) about the impacts of congestion pricing in Manhattan. Four different charging schemas were compared. We implement the highest charging schema from that plan in our virtual test bed (Schema 1). Another schema (Schema 2) is tested with $14 per passenger vehicle according to a new pilot program (Holland and Shah, 2019). The simulated charging schemas are presented in Table 6. All prices are charged two-way (upon entry and exit), which is consistent with RPA's report.

**Table 6. Simulated charging schemas of congestion pricing**

| Schema | 1 | 2 |
|---|---|---|
| Peak (6-10AM, 2-8PM) | $9.18 | $14 |
| Off-peak (5-6AM, 10AM-2PM, 8-11PM) | $3.06 | $3.06 |
| Night (11PM-5AM) | $3.06 | $3.06 |

## 5.2. Results

The congestion pricing was implemented using MATSim's extension "Road Pricing". By defining the boundary of the charging area, this extension can charge a fixed price on each car trip inside/outside the charge area. Different prices can be charged at different time periods. The congestion price charged per trip is calculated as a disutility (negative score) that is incorporated into the score of the agent's daily plan. The simulations were run on a desktop with Intel(R) Xeon(R) CPU E5-2637 v4 @ 3.50GHz processors and 128GB RAM. Each implementation of a schema ran with 100 iterations in MATSim and was initiated using the final plan set obtained from the base MATSim-NYC scenario. The computation time is approximately 15 hours per scenario.

Considering the charging area is not the whole Manhattan, we use a different population segment definition to evaluate the impact of the policies. Two new segments are defined as "Charging-related" vs. "Non-charging-related". A person who has at least one trip origin or destination in the charging area belongs to the Charging-related segment (including travelers making trips completely within the charging area without paying a cordon fee); otherwise, they belong to the Non-charging-related segment.

One analysis that is possible due to the use of an agent-based simulation is the tracking of mode shifts from travelers due to congestion pricing. The shift in trips in the Charging-related segment are shown in Figure 17. A significant drop in car trips is found under both schemas, where the higher price leads to a larger drop. Compared to the 59,000 decrease of daily car trips from RPA, our simulation indicates a 127,000-trip decline under the same schema (Schema 1). Trips of all modes except for transit and walk decrease after charging the congestion price. These outcomes are in line with expectations since Manhattan becomes less congested and more people would use transit instead of car.

The daily and annual revenues collected under each schema are calculated and presented in Table 7. The daily revenues are obtained from the simulation and expanded to annual revenues assuming 261 equivalent working days, with 4% legislated exemptions, $113M operating cost, and $30M deduction from a technical memorandum (RPA, 2019). The results are consistent with RPA's projection, although our simulation reports double the reduction in cars shifted to other modes.

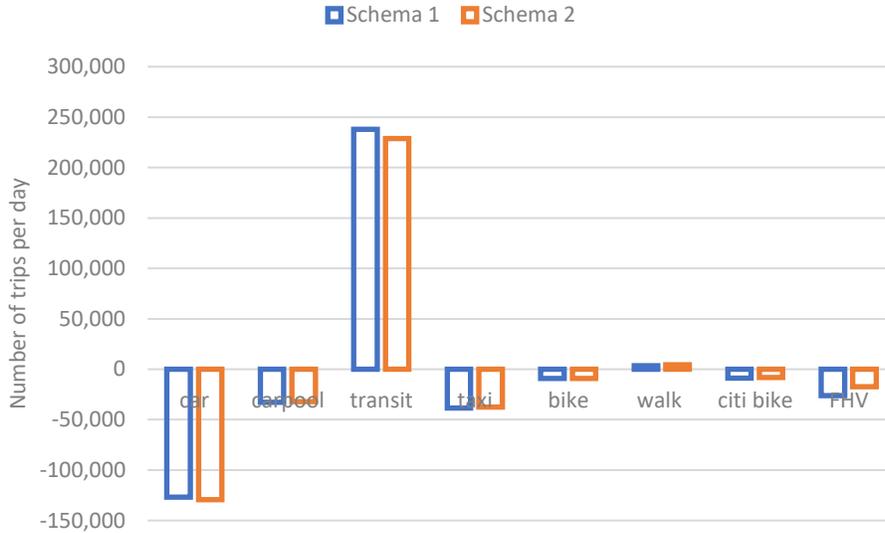

**Figure 17. Daily trip shifts after charging congestion price for Charging-related segment.**

**Table 7. Revenues collected under each schema**

|  | Forecasted revenues using MATSim-NYC | | RPA's proposal |
|---|---|---|---|
|  | **Schema 1** | **Schema 2** | **Schema 1** |
| **Daily revenue** | 4.91M | 6.75M | N/A |
| **Annual** | 1.09B | 1.55B | 1.09B |

The changes in realized consumer surplus for travelers relative to the baseline scenario are estimated and presented in Table 8. The realized consumer surplus is calculated as the executed daily plan score on the final day of the simulation, less the activity participation portion (Eq. (7)) and the late arrival portion (Eq. (8)) as discussed in Section 4.1.2. As such, the values in Table 8 reflect changes in only the travel portion of the consumer surplus, which includes the disutility of the congestion pricing payment. The marginal utility of cost is based on $\beta_{cost} = -0.06$ from He et al. (2020), which corresponds to a VOT of $29/h (originally estimated for Manhattan segment but assumed for all of NYC as a conservative value because non-Manhattan segment cost is statistically insignificant).

While the averages in Table 8 counter-intuitively suggest that everyone in the population gains a surplus from congestion pricing even considering the congestion charge, that is not the case. The cumulative distribution of changes in daily consumer surplus relative to base scenario under Schema 1 is shown in Figure 18. In the Charging-related segment, 37.3% of the population suffer reductions in travel consumer surplus due to congestion pricing. In the Non-charging-related segment, the proportion is slightly higher at 39.9%. The figure highlights one of the key advantages of using a multiagent simulation: it can distinguish the heterogeneous effects of congestion pricing on the population. Due to this heterogeneity, it is possible that a congestion charge can improve average consumer surplus even before the revenues are redistributed back to the population. Two other insights are gained.

**Table 8. Average change in daily travel-based consumer surplus ($) per capita by segment**

|  | Charging-related | Non-charging-related | Citywide |
|---|---|---|---|
| Schema 1 | +17.09 | +8.13 | +10.66 |
| Schema 2 | +17.11 | +7.79 | +10.55 |

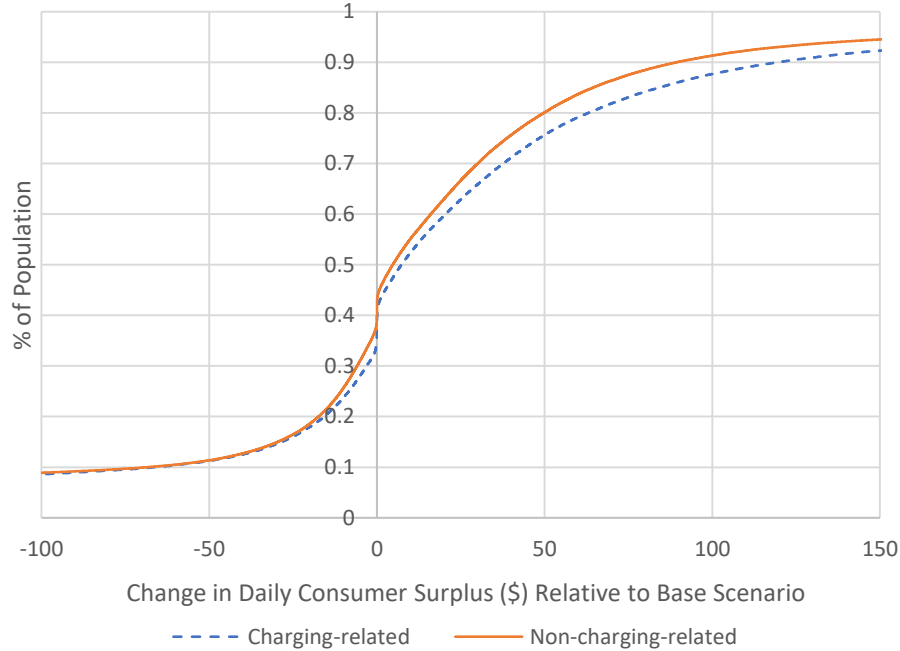

**Figure 18. Cumulative distribution of change in daily travel-based consumer surplus relative to base scenario under Schema 1.**

First, people from both population segments (Charging-related and Non-charging-related) have a net increase in consumer surplus (including the negative effects of the congestion payment), which indicates a positive impact on people in NYC from congestion pricing. People from the Charging-related segment (which include both Manhattan residents and those making trips into Manhattan) benefit more than those from the Non-charging-related segment, by an increase of 110 – 120%. This implies that the congestion pricing provides a net savings in travel times for the travelers within the charging area that significantly overcompensates for the cost of the charge on people crossing it. The result supports the congestion pricing policy around Manhattan. Furthermore, the results suggest that redistribution of toll revenues should focus on outer boroughs transit service to balance out the benefit.

Second, we see that Schema 2 results in a *lower* net benefit citywide than Schema 1 (+$10.55 vs +$10.66). This implies that the $14 charge may already be too high compared to the $9.18 charge, although it would further (incrementally) benefit folks in the charging area (+$17.11 vs +$17.09). This result suggests the importance of using policy tools like MATSim-NYC to help design the congestion pricing policy as it can offset benefits from one population segment against another. This finding would not have been possible using a static network policy tool.

The shifts in departure time before and after congestion pricing can be determined per mode. The departure time distributions of car for the two population segments are presented in Figure 19. The number of car trips in Charging-related segment reduced significantly across the whole day, especially in the peak periods (6 – 10 AM, 2 – 8 PM) when the congestion price is high. The distribution of car trips become overall flatter after charging the congestion pricing. Interestingly, the number of Non-charging-related segment car trips increase primarily in the off-peak periods, perhaps due to the gap in traffic left by the drop in the Charging-related segment benefiting other parts of the city.

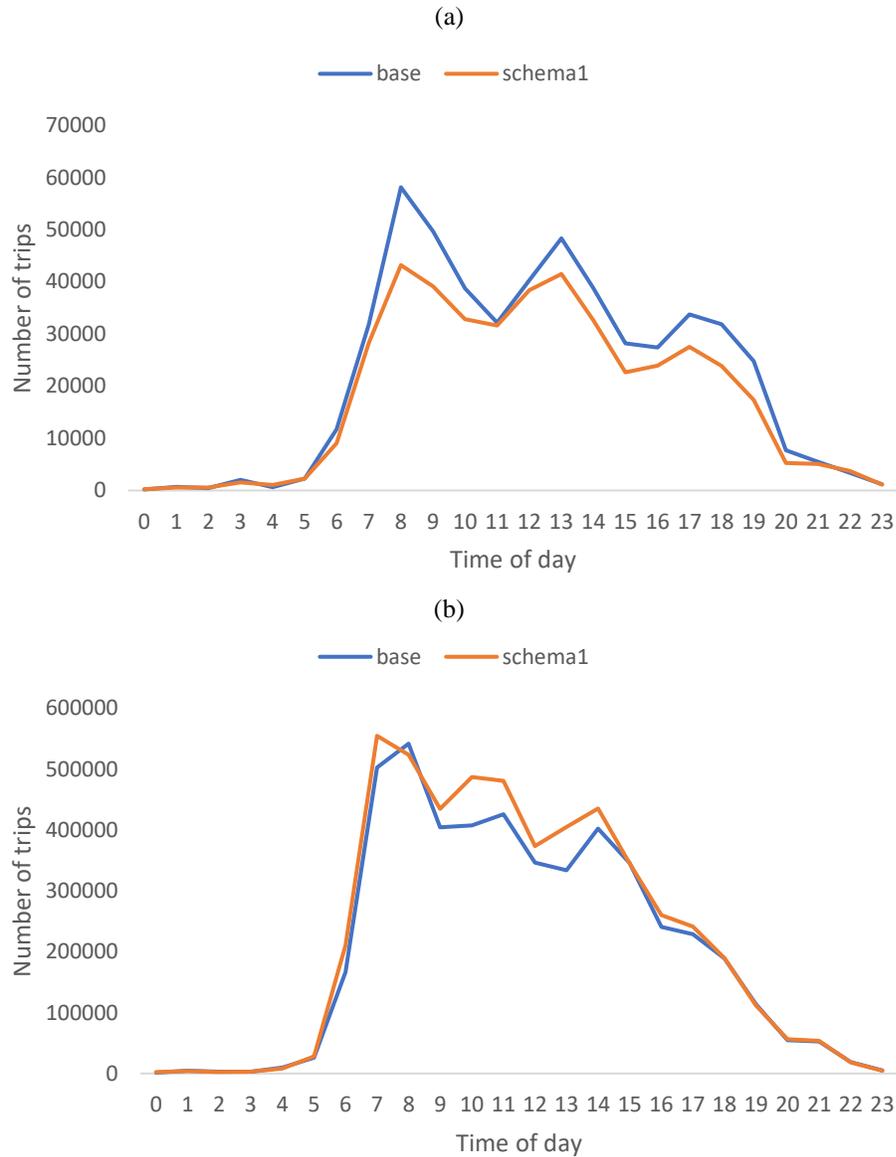

**Figure 19. Departure time distribution of car trips in (a) Charging-related segment (b) Non-charging-related segment.**

## 6. Conclusion

The impacts of city-scale policies on large cities like NYC are difficult to assess because of the heterogeneous population and complex transportation system. Existing regional travel demand models for NYC do not capture traffic dynamics or substitution effects of multiple modes that include FHV and bikeshare. Our study addresses this research gap with a new model called MATSim-NYC designed to evaluate emerging transportation technologies and policies. A multi-agent simulation testbed based on MATSim was developed for NYC and calibrated for a 4% sample population. Both the link unsaturated flow speed and capacity of road network were adjusted for six time periods and two road types. The calibrated model was validated with subway ridership data and key road corridor traffic count data. The predicted subway daily ridership of ten high-volume stations across the city has an 8% average relative error. The average relative difference of the traffic volumes is 39.8% while the median is 29%, which is a relatively well-behaved result in terms of a citywide simulation.

The calibrated simulation testbed is then used to evaluate the impacts of cordon-based congestion pricing policies in NYC, using a value of time of $29/h estimated from the synthetic population from He et al. (2020). One schema from RPA's proposal is implemented along with another pilot schema. The impacts of congestion pricing are evaluated for different segments and times of the day. Under the same schema, the car trip decrease is twice that of RPA's proposal, even while the annual revenues collected are similar.

More insights can be extracted from the testbed that can only be determined using the multiagent simulation. The first insight involves the change in consumer surplus. Although 37% to 40% of the population suffer negatively from the congestion cost, the benefits for the rest of the population far outweigh them resulting in average improvements to consumer surplus for both population segments. The result supports the congestion pricing policy. This finding is only possible because the multiagent simulation can model the heterogeneous impacts of the congestion pricing.

The Charging-related segment is shown to benefit more than twice that of the other segment, which has implications for the distribution of the toll revenues in favor of transit in the outer boroughs.

When the price is increased from Schema 1 to Schema 2, the citywide change of consumer surplus decreases a little, which suggests that $14 both ways is too high of a fee. However, the benefits for the Charging-related segment continue to go up, which suggests that the higher price benefits the Manhattan population at the expense of the overall city.

Another insight is the impact on people's departure times. The number of car trips drop significantly throughout the day for the Charging-related segment while there is a mix of impacts for the non-Charging-related segment.

In future research, we plan to incorporate dynamic congestion pricing into the baseline model by the "decongestion" contribution in MATSim (Kaddoura and Nagel, 2019). Congestion price can be determined by either position in the queue or congestion level to reduce traffic congestion. This new feature would provide more flexible insights to the decision-makers to help them improve the transportation system of the city. Impacts of other policies and emerging technologies like cargo bikes deployment, on-demand autonomous taxis, etc., can also be analyzed with our simulation model in the future. The marginal utility of cost does not currently take account of the individual or their household's income level, which can be addressed in future research.

**Acknowledgments**

This research was conducted with funding support from a C2SMART project, USDOT award #69A3551747124. The development of simulation test bed in MATSim involved collaboration with Dr. Balac Milos from ETH Zurich and Joon Park from New York City Department of Transportation. Dr. Abdullah Kurkcu and Yubin Shen supported the data sharing and management. Qi Liu assisted the calibration of transit vehicle capacity, Haoran Su helped with the preparation of transit schedules, and Reuben Juster helped with the data collection.## References

Auld, J., Hope, M., Ley, H., Sokolov, V., Xu, B., & Zhang, K. (2016). POLARIS: Agent-based modeling framework development and implementation for integrated travel demand and network and operations simulations. *Transportation Research Part C: Emerging Technologies*, *64*, 101-116.

Axelrod, R., and L. Tesfatsion. (2006). A Guide for Newcomers to Agent-Based Modeling in the Social Sciences. In Handbook of Computational Economics, Vol. 2: Agent-Based Computational Economics (L. Tesfatsion and K. L. Judd, eds.). Handbooks in Economics Series, Elsevier/North-Holland, Amsterdam, Netherlands. 2006.

Axhausen, K. W., Zimmermann, A., Schönfelder, S., Rindsfüser, G., & Haupt, T. (2002). Observing the rhythms of daily life: A six-week travel diary. *Transportation*, *29*(2), 95-124.

Baghestani, A., Tayarani, M., Allahviranloo, M., & Gao, H. O. (2020). Evaluating the Traffic and Emissions Impacts of Congestion Pricing in New York City. *Sustainability*, *12*(9), 3655.

Balmer, M., Rieser, M., Meister, K., Charypar, D., Lefebvre, N., & Nagel, K. (2009). MATSim-T: Architecture and simulation. In A. L. C. Bazzan & F. Klugl (Eds.), Multi-agent systems for traffic and transportation engineering (pp. 57–78). Hershey, PA: IGI Global.

Balmer, M., Meister, K., Rieser, M., Nagel, K., & Axhausen, K. W. (2008). Agent-based simulation of travel demand: Structure and computational performance of MATSim-T. Arbeitsberichte Verkehrs-und Raumplanung, 504.

Becker, H., Balac, M., Ciari, F., & Axhausen, K. W. (2020). Assessing the welfare impacts of shared mobility and Mobility as a Service (MaaS). *Transportation Research Part A: Policy and Practice*, *131*, 228-243.

Bernhardt, K. (2007). Agent-based modeling in transportation. Artificial Intelligence in Transportation, 72.

Bonabeau, E. (2002). Agent-Based Modeling: Methods and Techniques for Simulating Human Systems. PNAS, Vol. 99, No. 3, 2002a.

Bonabeau, E. (2002) Predicting the Unpredictable. Harvard Business Review. March 2002b, pp. 109–116.

BQX. Retrieved August, 2019 from http://www.bqx.nyc.

BQX. (2018). BQX Completion of Conceptual Design Report. Retrieved August, 2019 from http://www.bqx.nyc/wp-content/uploads/2018/09/BQX_Completion_of_Conceptual_Design_Report_2018.08.pdf.

Bradley, M., Bowman, J. L., & Griesenbeck, B. (2010). SACSIM: An applied activity-based model system with fine-level spatial and temporal resolution. *Journal of Choice Modelling*, 3(1), 5–31.

Brownstone, D., Ghosh, A., Golob, T. F., Kazimi, C., & Van Amelsfort, D. (2003). Drivers' willingness-to-pay to reduce travel time: evidence from the San Diego I-15 congestion pricing project. *Transportation Research Part A: Policy and Practice*, *37*(4), 373-387.

Burris, M., & Swenson, C. (1998). Planning Lee County's Variable-Pricing Program. *Transportation research record*, *1617*(1), 64-68.

Certicky, M., Jakob, M., Pibil, R., & Moler, Z. (2014). Agent-based simulation testbed for on-demand mobility services. Paper presented at the proceedings of the 3rd international workshop on agent-based mobility, traffic and transportation models, methodologies and applications (ABMTRANS), Hasselt, Belgium.

Cetin, N., Burri, A., & Nagel, K. (2003). A parallel queue model approach to traffic microsimulations. Proc. Transportation Research Board 82nd Annual Meeting, Washington, DC.

Chow, J. Y. J. (2018). *Informed Urban transport systems: Classic and emerging mobility methods toward smart cities*. Elsevier.

Chow, J. Y. J., & Djavadian, S. (2015). Activity-based market equilibrium for capacitated multimodal transport systems. *Transportation Research Part C: Emerging Technologies*, *59*, 2-18.

Chow, J. Y. J., Ma, Z., Lee, M., and Goldwyn, E. (2020c). Evaluation of bus redesign alternatives in transit deserts under ride-hail presence. C2SMART Project Report.

Chow, J. Y. J., Ozbay, K., He, B. Y., Zhou, J., Ma, Z., Lee, M., Wang, D., and Sha, D. (2020b). Multi-agent simulation-based virtual test bed ecosystem: MATSim-NYC. C2SMART Project Report.

Chow, J. Y. J., Rath, S., Yoon, G., Scalise, P., Alanis Saenz, S. (2020a). Spectrum of Public Transit Operations: from Fixed Route to Microtransit. FTA Report, NY-2019-069-01-00.

Chow, J. Y. J., & Recker, W. W. (2012). Inverse optimization with endogenous arrival time constraints to calibrate the household activity pattern problem. *Transportation Research Part B: Methodological*, *46*(3), 463-479.29

Chow, J. Y. J., & Regan, A. C. (2014). A surrogate-based multiobjective metaheuristic and network degradation simulation model for robust toll pricing. *Optimization and Engineering*, *15*(1), 137-165.
Ciari, F., Balac, M., & Axhausen, K. W. (2016). Modeling carsharing with the agent-based simulation MATSim: State of the art, applications, and future developments. *Transportation Research Record*, *2564*(1), 14-20.
Cich, G., Knapen, L., Maciejewski, M., Bellemans, T., & Janssens, D. (2017). Modeling demand responsive transport using SARL and MATSim. *Procedia Computer Science*, *109*, 1074-1079.
Dennis, J. E., and D. J. Schnabel. (1989). A view of unconstrained optimization. Handbooks in Operation Research and Management Science (G.L. Nemhauser et al., Eds), Vol. 1, 1989, pp. 1-72.
Dia, H. (2002). An Agent-Based Approach to Modeling Driver Route Choice Behavior Under the Influence of Real-Time Information. Transportation Research Part C, Vol. 10, 2002, pp. 331–349.
Djavadian, S., & Chow, J. Y. J. (2017a). Agent-based day-to-day adjustment process to evaluate dynamic flexible transport service policies. *Transportmetrica B: Transport Dynamics*, *5*(3), 281-306.
Djavadian, S., & Chow, J. Y. J. (2017b). An agent-based day-to-day adjustment process for modeling 'Mobility as a Service' with a two-sided flexible transport market. *Transportation research part B: methodological*, *104*, 36-57.
Electric Railroad Association, (2016). The Bulletinl Vol. 59, No. 7. Retrieved from http://web.mta.info/mta/compliance/pdf/Title-VI-NYCT-Bus-Policies.pdf
Erath, A and Chakirov, A. 2016. Singapore. In: Horni, A, Nagel, K and Axhausen, K W. (eds.) The Multi-Agent Transport Simulation MATSim, Pp. 379–382. London: Ubiquity Press.
FHWA, Congestion Pricing: Examples Around the US. Retrieved December 2019 from https://ops.fhwa.dot.gov/congestionpricing/assets/us_examples.pdf.
FHWA, Congestion Pricing A Primer: Overview. Retrieved December 2019 from https://ops.fhwa.dot.gov/publications/fhwahop08039/fhwahop08039.pdf.
Flyvbjerg, B., Skamris Holm, M. K., & Buhl, S. L. (2005). How (in) accurate are demand forecasts in public works projects?: The case of transportation. *Journal of the American planning association*, *71*(2), 131-146.
Goulias, K. G., Bhat, C. R., Pendyala, R. M., Chen, Y., Paleti, R., Londuri, K. C., et al. (2011). Simulator of activities, greenhouse emissions, networks, and travel (SimAGENT) in southern California. Paper presented at the Transportation Research Board 91st Annual Meeting, Washington, DC.
Haglund, N., Mladenović, M. N., Kujala, R., Weckström, C., & Saramäki, J. (2019). Where did Kutsuplus drive us? Ex post evaluation of on-demand micro-transit pilot in the Helsinki capital region. *Research in Transportation Business & Management*, 100390.
He, B. Y. (2020). Code for MATSim-NYC project, Zenodo, doi: 10.5281/zenodo.4082007.
He, B. Y., Zhou, J., Ma, Z., Chow, J. Y. J., and Ozbay, K. (2020). Evaluation of city-scale built environment policies in New York City using an emerging mobility-accessible synthetic population. *Transportation Research Part A* 141, 444-467.
Hidas, P. (2002). Modeling Lane Changing and Merging in Microscopic Traffic Simulation. Transportation Research Part C, Vol. 10, 2002, pp. 351–371.
Hörl, S., Balac, M., & Axhausen, K. W. (2019, June). Dynamic demand estimation for an AMoD system in Paris. In 2019 IEEE Intelligent Vehicles Symposium (IV) (pp. 260-266). IEEE.
Hörl, S., Ruch, C., Becker, F., Frazzoli, E., & Axhausen, K. W. (2019). Fleet operational policies for automated mobility: A simulation assessment for Zurich. *Transportation Research Part C: Emerging Technologies*, *102*, 20-31.
Horni, A, Nagel, K and Axhausen, K W. (2016). Introducing MATSim. In: Horni, A, Nagel, K and Axhausen, K W. (eds.) The Multi-Agent Transport Simulation MATSim, Pp. 3–8. London: Ubiquity Press. DOI: http://dx.doi.org/10.5334/baw.1. License: CC-BY 4.0.
Horni, A and Nagel, K. (2016). More About Configuring MATSim. In: Horni, A, Nagel, K and Axhausen, K W. (eds.) The Multi-Agent Transport Simulation MATSim, Pp. 35–44. London: Ubiquity Press. DOI: http://dx.doi.org/10.5334/baw.4. License: CC-BY 4.0
JOSM. (2018). Available at https://josm.openstreetmap.de.
Kaddoura, I., & Nagel, K. (2019). Congestion pricing in a real-world oriented agent-based simulation context. *Research in Transportation Economics*, 74, 40-51.
Macal, C. M., & North, M. J. (2006b). Tutorial on agent-based modeling and simulation Part 2: How to model with agents. Proceedings of the 2006 Winter Simulation Conference, Monterey, CA.
MATSim (2020). Retrieved December 01, 2020, from https://github.com/matsim-scenarios.
Metropolitan Transportation Authority, General Transit Feed Specification. Accessed in August 2018 at https://transitfeeds.com/p/mta.
Metropolitan Transportation Authority, (2018a). MTA New York City Transit and MTA Bus Company System-wide

Service Standards. Retrieved from: http://web.mta.info/mta/compliance/pdf/Title-VI-NYCT-Bus-Policies.pdf.

Metropolitan Transportation Authority, (2018b). Average Weekday Subway Ridership. Retrieved from: http://web.mta.info/nyct/facts/ridership/ridership_sub.html.

Metropolitan Transportation Authority, (2018c). Introduction to Subway Ridership. Retrieved from: http://web.mta.info/nyct/facts/ridership/

Nagel, K., Beckman, R. L., & Barrett, C. L. (1999). TRANSIMS for transportation planning. Paper presented at the 6th International Conference on Computers in Urban Planning and Urban Management, Franco Angeli, Milano, Italy.

Nagel, K, Kickhöfer, B, Horni, A and Charypar, D. (2016). A Closer Look at Scoring. In: Horni, A, Nagel, K and Axhausen, K W. (eds.) The Multi-Agent Transport Simulation MATSim, Pp. 23–34. London: Ubiquity Press. DOI: http://dx.doi.org/10.5334/baw.3. License: CC-BY 4.0

Nahmias-Biran, B.H., Oke, J.B., Kumar, N., Basak, K., Araldo, A., Seshadri, R., Akkinepally, A., Lima Azevedo, C. and Ben-Akiva, M., 2019. From Traditional to Automated Mobility on Demand: A Comprehensive Framework for Modeling On-Demand Services in SimMobility. *Transportation Research Record*, p.0361198119853553.

Neumann, A. 2016. Berlin I: BVG Scenario. In: Horni, A, Nagel, K and Axhausen, K W. (eds.) The Multi-Agent Transport Simulation MATSim, Pp. 369–370. London: Ubiquity Press. DOI: http://dx.doi.org/10.5334/ baw.53. License: CC-BY 4.0

New York Metropolitan Transportation Council, (2000). Traffic Analysis Zone shapefiles, 2010/2011 Regional Household Travel Survey, available by request.

New York City Department of Transportation. Best Practice Model. Retrieved December 2019 from https://www1.nyc.gov/html/dot/downloads/pdf/cigsdts_best_practice_model.pdf.

New York City Department of Transportation. Citywide Mobility Survey. http://www.nyc.gov/html/dot/downloads/pdf/nycdot-citywide-mobility-survey-report-2017.pdf. Accessed August, 2018.

New York City Department of Transportation. (2016). 2016 New York City Bridges Traffic Volumes. Retrieved December 2019 from http://www.nyc.gov/html/dot/downloads/pdf/nyc-bridge-traffic-report-2016.pdf.

Rieser, M., Dobler, C., Dubernet, T., Grether, D., Horni, A., Lammel, G., Wariach, R., Zilske, M., Axhausen, K.W. & Nagel, K. (2014). MATSim user guide. Zurich: MATSim.

Rieser, M, Horni, A and Nagel, K. (2016). Scenarios Overview. In: Horni, A, Nagel, K and Axhausen, K W. (eds.) The Multi-Agent Transport Simulation MATSim, Pp. 367–368. London: Ubiquity Press.

Rieser-Schüssler, N, Bösch, P M, Horni, A and Balmer, M. 2016. Zürich. In: Horni, A, Nagel, K and Axhausen, K W. (eds.) The Multi-Agent Transport Simulation MATSim, Pp. 375–378. London: Ubiquity Press.

Robbins, C. (Jan, 10th, 2020). De Blasio's Still Pushing The $2.7 Billion Streetcar Plan You Probably Forgot About. The Gothamist. Retrieved from https://gothamist.com/news/bqx-streetcar-nyc-to-shelbyville.

Ronald, N., Thompson, R. & Winter, S. (2015). Simulating Demand-responsive Transportation: A Review of Agent-based Approaches, Transport Reviews, 35:4, 404-421, DOI: 10.1080/01441647.2015.1017749.

Rothfeld, R., Balac, M., Ploetner, K. O., & Antoniou, C. (2018). Agent-based simulation of urban air mobility. In *2018 Modeling and Simulation Technologies Conference* (p. 3891).

RPA (2019). Congestion Pricing in NYC: Getting it right. Retrieved November, 2019 from https://www.rpa.org/publication/congestion-pricing-in-nyc-getting-it-right.

Schaller, B. (2010). "New York City's congestion pricing experience and implications for road pricing acceptance in the United States," Transport Policy, Elsevier, vol. 17(4), pages 266-273, August.

Small, K. (1982). The Scheduling of Consumer Activities: Work Trips. The American Economic Review, 72(3), 467-479. Retrieved June 2, 2020, from www.jstor.org/stable/1831545

Small, K. A. (1992). Using the revenues from congestion pricing. *Transportation*, *19*(4), 359-381.

Spall, J. C. (1988). A stochastic approximation algorithm for large-dimensional systems in the Kiefer-Wolfowitz setting. Presented at the 27th Conference on Decision and Control, Austin, Texas, USA, 1988.

Spall, J. C. (1998). Implementation of the simultaneous perturbation algorithm for stochastic optimization. IEEE Transactions on Aerospace and Electronic Systems, Vol. 34, No. 3, 1998a, pp. 817-823.

Spall, J. C. (1998). An overview of the simultaneous perturbation method for efficient optimization. Johns Hopkins APL Technical Digest, Vol. 19, 1998b, pp. 482-492.

Von Neumann, J. (1966). *Theory of Self-Reproducing Automata*. Edited by A. W. Burk. Urbana: University of Illinois Press.

Wang, D., He, B. Y., Gao, J., Chow, J. Y., Ozbay, K., & Iyer, S. (2020). Impact of COVID-19 Behavioral Inertia on Reopening Strategies for New York City Transit. *arXiv preprint arXiv:2006.13368*.

Wong, Y. Z., Hensher, D. A., & Mulley, C. (2020). Mobility as a service (MaaS): Charting a future

context. *Transportation Research Part A: Policy and Practice*, *131*, 5-19.

Yang, H., & Huang, H. J. (1998). Principle of marginal-cost pricing: how does it work in a general road network?. *Transportation Research Part A: Policy and Practice*, *32*(1), 45-54.

Ye, X., Konduri, K., Pendyala, R. M., Sana, B., & Waddell, P. (2009). A methodology to match distributions of both household and person attributes in the generation of synthetic populations. In 88th Annual Meeting of the Transportation Research Board, Washington, DC.

Yu, M., and W. D. Fan. (2017). Calibration of microscopic traffic simulation models using metaheuristic algorithms. International Journal of Transportation Science and Technology, Vol. 6, 2017, pp. 63-77.

Zhang, L. (2006). An Agent-Based Behavioral Model of Spatial Learning and Route Choice. Presented at 85th Annual Meeting of the Transportation Research Board, Washington, D.C., 2006.

Zhang, X., & Yang, H. (2004). The optimal cordon-based network congestion pricing problem. *Transportation Research Part B: Methodological*, *38*(6), 517-537.

Zheng, H., Son, Y. J., Chiu, Y. C., Head, L., Feng, Y., Xi, H., Kim, S., & Hickman, M. (2013). A primer for agent-based simulation and modeling in transportation applications (No. FHWA-HRT-13-054). United States. Federal Highway Administration.

Ziemke, D. 2016. Berlin II: CEMDAP-MATSim-Cadyts Scenario. In: Horni, A, Nagel, K and Axhausen, K W. (eds.) The Multi-Agent Transport Simulation MATSim, Pp. 371–372. London: Ubiquity Press. DOI: http://dx.doi.org/10.5334/baw.54. License: CC-BY 4.0

Ziemke, D., Kaddoura, I., & Nagel, K. (2019). The MATSim Open Berlin Scenario: A multimodal agent-based transport simulation scenario based on synthetic demand modeling and open data. *Procedia computer science*, *151*, 870-877.